\documentclass[11pt,a4paper]{article}
\pdfoutput=1
\usepackage{jheppub}
\usepackage{gensymb}
\usepackage{subfigure}
\usepackage{amssymb,amsmath}
\usepackage{graphicx}
\usepackage{color}
\usepackage{cancel}
\usepackage[colorlinks=true
,urlcolor=blue
,citecolor=blue
,linkcolor=blue
,pagecolor=blue
,linktocpage=true
,pdfproducer=medialab
]{hyperref}
\usepackage[section]{placeins}
 \usepackage[numbers]{natbib}
\usepackage{notoccite}
\makeatletter \renewcommand{\@dotsep}{10000} \makeatother
\def\be{\begin{equation}}
\def\ee{\end{equation}}
\def\bea{\begin{eqnarray}}
\def\eea{\end{eqnarray}}
\def\bi{\begin{itemize}}
\def\ei{\end{itemize}}

\usepackage[nodisplayskipstretch]{setspace}

\def\mgut{M_{\rm GUT}}

\newcommand{\beq}{\begin{equation}}
\newcommand{\eeq}{\end{equation}}

\newcommand*{\Scale}[2][4]{\scalebox{#1}{$#2$}}%
\begin{document}

\begin{titlepage}
\pagestyle{empty}

\vspace*{0.2in}
\begin{center}
{\Large \bf Mono-Photon Events with Light Higgs bosons in Secluded UMSSM} \\
\vspace{1cm}
{\bf  Ya\c{s}ar Hi\c{c}y\i lmaz$^{a,b,}$\footnote{Email: yasarhicyilmaz@balikesir.edu.tr; Y.Hicyilmaz@soton.ac.uk},
Levent Selbuz$^{c,}$\footnote{Email: selbuz@eng.ankara.edu.tr} {\rm and}
Cem Salih $\ddot{\rm U}$n$^{d,}\hspace{0.05cm}$\footnote{E-mail: cemsalihun@uludag.edu.tr}}
\vspace{0.5cm}

{\it
$^a$Department of Physics, Bal\i kesir University, TR10145, Bal\i kesir, Turkey, \\
$^b$School of Physics and Astronomy, University of Southampton, Highfield, Southampton SO17 1BJ, United Kingdom\\
$^c$Department of Engineering Physics, Ankara
University, TR06100 Ankara, Turkey, \\
$^d$Department of Physics, Bursa Uluda\~{g} University, TR16059 Bursa, Turkey, \\
}
\end{center}

\vspace{0.1cm}
\begin{abstract}

We explore the Higgs boson implications of a class of supersymmetric models, which extends the MSSM gauge group by a $U(1)^{\prime}$ symmetry, which is broken at low energy scales by VEVs of four MSSM singlet fields. These singlets also form a secluded sector, and one of them is allowed to interact with the MSSM Higgs fields directly. After the $U(1)^{\prime}$ and electroweak symmetry breaking, the low scale spectra include six CP-even and 4 CP-odd Higgs bosons, whose masses can lie from 80 GeV to 2-3 TeV. We find that the heavy CP-even Higgs bosons can be probed through their decays into a pair of SM gauge bosons currently up to about $m_{h_{i}} \simeq 1.5$ TeV, while their probe can be extended to about $m_{h_{i}} \simeq 2.5$ TeV in near future. The most interesting feature of the low scale spectra in the class of Secluded $U(1)^{\prime}$ models is to include two light CP-odd Higgs bosons whose masses are bounded at about 250 GeV by the current collider and DM experiments, when the LSP neutralino is mostly formed by the MSSM singlets. These light CP-odd Higgs bosons should be formed by the MSSM singlet scalars to be consistent with the current constraints. Despite their singlet nature, they can be traced through their associated production with photons. In our work, we consider their productions at the collider experiments together with photons, and we realize that these light Higgs bosons can potentially be probed during Run-3 experiments of LHC when they are lighter than about 100 GeV. We also show that Run-4 and HL-LHC experiments will be able to probe these light scalars up to about 250 GeV.

\end{abstract}

\end{titlepage}

\tableofcontents

\noindent \hrulefill

\section{Introduction}
\label{sec:intro}

Even though the Standard Model (SM) is undoubtedly an effective theory, which can describe the physcics precisely up to some energy levels at the order TeV scale, the absence of any direct signal for new physics beyond SM (BSM) leads to consider non-minimally constructed BSM models. {Despite the lack of a direct signal, these models can still be confronted indirectly with some experimental results such as those in semi-leptonic $B-$meson decays \cite{HFLAV:2022pwe,CMS:2020rox,Belle-II:2022hys}, lepton anomalous magnetic moments \cite{Muong-2:2021ojo,Hanneke:2010au,Morel:2020dww} etc., which measure some deviations from the SM predictions.} These deviations also potentially indicate an enhanced non-universality in the leptons, which may necessitate to consider BSM models non-trivially distinguishing the flavors \cite{Zhu:2021vlz,Langacker:2000ju,Barger:2003hg,Badziak:2019gaf,Boubaa:2022xsk,Frank:2021nkq}. In addition, the searches for the extra Higgs bosons also take important place to probe the BSM models. The current analyses can exclude the extra Higgs bosons up to about 2 TeV if such Higgs bosons are allowed to decay into a pair of $\tau-$leptons \cite{ATLAS:2020zms}. There have also been some excesses reported which might accommodate new scalars at low scale mass scales such as around 30 GeV \cite{CMS:2018lce}, 90 GeV \cite{CMS:2018cyk}, 130 GeV \cite{ATLAS:2021zyv} etc. These searches are also being strengthened by the analyses considering the associated production of Higgs bosons \cite{ATLAS:2021gcn,CMS:2020krr}. The associated productions of a Higgs boson can play a crucial role especially when it is a singlet under the SM gauge group and/or it is connected to a dark matter (DM) candidate \cite{ATLAS:2018yax,CMS:2018fux,ATLAS:2018jsf,ATLAS:2018fcn}.

If there happens any observation and/or deviation from the expected results in these Higgs searches, it definitely indicates a new state which can be accommodated in BSM models. Formulating the BSM physics, among many others, supersymmetry (SUSY) is one of the forefront candidate. The main motivation behind the SUSY models arises from stabilizing the Higgs mass against the quadratic divergencies \cite{Gildener:1976ai,Gildener:1979dd,Weinberg:1978ym,Susskind:1978ms,Veltman:1980mj} and the scalar potential \cite{Degrassi:2012ry,Bezrukov:2012sa,Buttazzo:2013uya,Branchina:2013jra,Branchina:2014usa,R.:2019ply}. In addition, imposing $R-$parity conservation to avoid the fast proton decay leads to pleasant DM candidates whose implications can be tested in the current experiments. However, the minimal supersymmetric extension of SM (MSSM), the extra Higgs bosons receive a strong negative impact from the current experiments. Apart from the analyses over their $\tau\tau$ decay modes, the current constraints on rare $B-$meson decays such as $B_{s}\rightarrow \mu^{+}\mu^{-}$ and $B\rightarrow X_{s}\gamma$ themselves can exclude the solutions when $m_{H,A}\lesssim 400-500$ GeV \cite{Babu:2020ncc,Gomez:2020gav,Raza:2018jnh}, and hence, they might be subjected to the heavy Higgs boson searches. On the other hand, a slight step away from MSSM by adding a singlet and/or supplementing the MSSM gauge group with new symmetries can accommodate light scalar states in the mass spectrum, which are preferably singlets. One of the simplest extension of MSSM referred to next to MSSM (NMSSM) involves a singlet superfield ($\widehat{S}$), whose scalar component allows to develop a non-zero vacuum expectation value (VEV) and interact with the MSSM Higgs fields at tree-level. It also yields a non-trivial mixing between this singlet state and the MSSM Higgs bosons, and possibly yield some tracks at the low scales \cite{Dedes:2000jp,Stal:2011cz,Gabelmann:2019jvz} and through its radiative contributions to the SM-like Higgs boson mass it loosens the requirement of the heavy SUSY particles and/or large trilinear couplings (see for instance, \cite{Hicyilmaz:2017nzo,Hicyilmaz:2017ntm}). However, the NMSSM framework might become more complicated to be simplistic extension of MSSM due to the existence of massless Goldstone boson, since it is strongly constrained by the current cosmological observations \cite{ParticleDataGroup:2002ivw}. Even though the massless states can be avoided by adding a self interaction term of the singlet field ($S^{3}$), then there arises the domain-wall problem (for detailed reviews see, \cite{Maniatis:2009re,Panagiotakopoulos:1998yw,Panagiotakopoulos:1999ah,Panagiotakopoulos:2000wp}).

The domain-wall problem can be avoided if the singlet field is connected to another $U(1)$ symmetry. These models form a class of $U(1)^{\prime}$ extended SUSY models (UMSSM), and they are also favored by the solution of the $\mu-$problem. When the MSSM fields are non-trivially charged under the $U(1)^{\prime}$ symmetry, the biliniear mixing term of the MSSM Higgs fields is effectively generated by the VEV of the Singlino field through $\lambda SH_{u}H_{d}$. In addition, UMSSM models extend the particle content further by adding $Z^{\prime}$ and its superpartner associated with the local $U(1)^{\prime}$ symmetry, the right-handed neutrinos to cancel the anomalies. One can also consider further extensions in the particle content. For instance, if there exists only one singlet field ($S$), then the heavy mass bounds on $Z^{\prime}$ \cite{Scutti:2021kjp,CMS:2019gwf,CMS:2021ctt} requires very large VEV for $S$ ($v_{S}$) leading to heavy $U(1)^{\prime}$ sector, which decouples from MSSM, and its low scale implications cannot be distinguished from the MSSM \cite{Hicyilmaz:2020bph,Frank:2020pui,Hicyilmaz:2017ntm}. Besides, the Higgs potential should be carefully analyzed, since there exist some directions of symmetry breaking in which either the VEVs of the MSSM Higgs fields vanish ($v_{u}\sim v_{d}\sim 0$) or $v_{S}$ happens to be comparable with $v_{u,d}$ ($v_{S}\sim v_{d}\sim v_{u}$). The former case definitely contradicts with the electroweak symmetry breaking, while the latter cancels the hierarchy between $Z^{\prime}$ and $Z$ \cite{Erler:2002pr}. These problems can be avoided by adding three more MSSM singlet fields, and the superpotential can be formed as follows: 

\begin{equation}
\setstretch{2.0}
\begin{array}{rll}
\widehat{W} & = W_{{\rm MSSM}}(\mu = 0) & + \lambda \widehat{S}\widehat{H}_u \cdot
        \widehat{H}_d + {h_{\nu}} \widehat{L}\cdot
        \widehat{H}_u \widehat{N}+
        \frac{\kappa}{3} \widehat{S}_1 \widehat{S}_2 \widehat{S}_3 \\
         && +\sum_{i=1}^{n_{\cal{Q}}} {h}_Q^i \widehat{S} \widehat{\cal{Q}}_i
        \widehat{\cal{\overline{Q}}}_i + \sum_{j=1}^{n_{\cal{L}}} {h}_L^j
        \widehat{S} \widehat{\cal{L}}_j \widehat{\cal{\overline{L}}}_j
\end{array}
\label{eq:superpot}
\end{equation}
where the MSSM superfields of quarks and leptons $\widehat{Q}, \widehat{U}, \widehat{D},\widehat{L}$ and $\widehat{E}$ are included in $W_{{\rm MSSM}}$, and $\widehat{H}_{u}, \widehat{H}_{d}$ denote the MSSM Higgs doublets. Note that we do not assume $Q^{\prime}_{H_{d}}+Q^{\prime}_{H_{u}}=0$ for the $U(1)^{\prime}$ charges of $H_{u}$ and $H_{d}$, and thus $\mu H_{u}H_{d}$ is not allowed by the gauge invariance, which is indicated in the argument of $W_{{\rm MSSM}}$. The new fields required by the $U(1)^{\prime}$ extension are assumed to be MSSM singlets, which are $S,S_{1,2,3}$ and the right-handed neutrino superfield $\widehat{N}$. We refer to this class of UMSSM models with three additional MSSM singlet fields ($S_{1,2,3}$) to the Secluded UMSSM \cite{Erler:2002pr,Chiang:2008ud,Demir:2010is,Frank:2012ne}. Despite the existence of the right-handed neutrinos, the anomaly cancellation in these models, in general, require also the exotic quarks $\mathcal{Q}_{i}$ and leptons $\mathcal{L}_{i}$. One of the simplest choices to satisfy the anomaly cancellation is to include a color triplet exotics with $Q_{\mathcal{Q}}=3, Y_{\mathcal{Q}}=-1/3$ and a color singlet with $Q_{\mathcal{L}}=2, Y_{\mathcal{L}}=-1$ \cite{Cheng:1998nb,Cheng:1998hc,Erler:2000wu,Langacker:2000ju,Barger:2003hg,Demir:2005ti,Hicyilmaz:2021oyd}. Note that these exotics are singlet under $SU(2)_{L}$ while their charges under $U(1)^{\prime}$ and $U(1)_{Y}$ are given as $Q$ and $Y$, 
respectively.

In the presence of the additional MSSM singlet fields, which are charged under the $U(1)^{\prime}$ group and allowed to develop VEVs, the heavy $Z^{\prime}$ mass can be realized even the $U(1)^{\prime}$ symmetry is broken at around TeV scale, since all the $U(1)^{\prime}$ breaking VEVs contribute. In this case, the Secluded UMSSM sector does not have to be decoupled from the MSSM sector, and it can significantly alter the low scale implications. Moreover, the models in this class can emerge from a larger symmetry which is broken at the grand unified scale ($\mgut$) such as $E_{6}$ through the following breaking chain:

\begin{equation}
\begin{array}{rl}
E_{6} & \rightarrow SO(10)\times U(1)_{\psi} \\
& \rightarrow SU(5)\times U(1)_{\chi}\times U(1)_{\psi} \\
& \rightarrow SU(3)_{C}\times SU(2)_{L}\times U(1)_{Y}\times U(1)^{\prime}
\end{array}
\label{eq:breaking}
\end{equation}

In general treatments, the resulting $U(1)^{\prime}$ group in the last line of Eq.(\ref{eq:breaking}) is considered as a linear combination of $U(1)_{\chi}$ and $U(1)_{\psi}$ with the charge $Q^{\prime} = Q_{\chi} \cos \theta_{E_6} + Q_{\psi} \sin \theta_{E_6}$. However, the models in the UMSSM class are not limited to be a combination of these two groups, and different charge assignments can be found as solutions of the anomaly cancellations \cite{Demir:2010is,Demir:2005ti,DelleRose:2018eic,Hicyilmaz:2021oyd,Frank:2021nkq}. A recent study \cite{Hicyilmaz:2021khy} (one of ours) has considered such a model which is constrained from $\mgut$ by imposing the universal soft supersymmetry breaking (SSB) masses for both the scalars and gauginos. The MSSM fields are assigned to family-independent $U(1)^{\prime}$ charges. The results for the DM implications within this class of models are represented for several distinct $U(1)^{\prime}$ charge assignments, and it has been shown that the MSSM singlets can play a crucial role by forming the LSP neutralino at the mass scale below about 200 GeV. Since the gaugino masses at $\mgut$ is assumed to be universal, the MSSM gauginos exhibit the Constrained MSSM (CMSSM) mass relations at the low scale ($M_{1}:M_{2}:M_{3}\simeq 1:2:4$ \cite{Bae:2016dxc}) {and the neutral gauginos - Bino and Wino - can form experimentally consistent DM solutions} of the LSP neutralino at about 600 GeV or more. In these mass scales, their masses are comparable with the Higgsino masses (due to lower values of $\mu-$term), thus the LSP happens to be a mixture of Bino-Higgsino. The presence of the Higgsinos in the LSP composition, on the other hand, yield large cross-section in the DM scatterings at nuclei, and thus they are excluded by the current bounds from the direct detection experiments \cite{Akerib:2018lyp,Aalbers:2016jon,Aprile:2020vtw,Tanaka:2011uf,Khachatryan:2014rra,Abbasi:2009uz,Akerib:2016lao}. Note that this results arise directly from the universal gaugino masses imposed at $\mgut$ in our analyses. In this case, the gluino mass bound ($m_{\tilde{g}}\gtrsim 2.1$ TeV) \cite{Aaboud:2017vwy} imposes an impact also on the Bino and Wino masses which bounds their masses at about 600 GeV from below. This tension can be removed if the non-universal gaugino masses at $\mgut$ is considered \cite{Gomez:2020gav,Shafi:2021jcg,Gomez:2022qrb}.

It has been observed that in the light DM solutions ($m_{\tilde{\chi}_{1}^{0}}\gtrsim 100$ GeV), the DM relic density can be saturated by the MSSM singlets. While the LSP neutralino can be purely Singlino when $m_{\tilde{\chi}_{1}^{0}}\lesssim 200$ GeV, Singlino can still determine the main nature of DM up to about $80\%$ for $200 \lesssim m_{\tilde{\chi}_{1}^{0}}\lesssim 500$ GeV. Even though $S_{1,2,3}$ interact only among themselves, $S$ can provide a weakly interacting DM particle through the Higgs portal, since it has a tree-level coupling to the MSSM Higgs fields. This connection between the Singlino and the MSSM Higgs fields leads to relatively large scattering cross-section which can be tested in XENON experiment in near future \cite{Aprile:2020vtw}. Apart from the direct detection experiments, the current relic density measurements by the Planck satellite \cite{Planck:2018vyg} also provide a strong constraint in the mass spectrum. The correct relic density of Singlino-like DM can be realized through the annihilations of two LSP into Higgs bosons, which are lighter than 300 GeV. The typical spectra of this class of UMSSM models involve 4 CP-odd Higgs bosons, and as mentioned before, since the MSSM Higgs bosons have a lower mass bound at about $400-500$ GeV due to several constraints, these light CP-odd Higgs bosons are required to be mostly singlet under the MSSM gauge group, though the MSSM Higgs field can be involved through the mixing. This mixing can be constrained further through the MSSM Higgs boson decays into these singlet Higgs bosons and/or their decays into the SM gauge bosons. Once a consistent mixing is obtained between the singlet and MSSM Higgs fields, such light Higgs bosons can be subjected to the mono $Z/\gamma$ signals and confronted with the current analyses. 

In our work, we will discuss the Secluded UMSSM models and possible signals that can probe/test the light Higgs bosons, which is briefly described above. Even though the low scale implications of the models in this class can yield richer implications, we assume that these models emerge from breaking of larger symmetries proposed by grand unified theories (GUTs) at $\mgut$. In its construction, apart from the different charge assignments, we follow its minimal construction by assuming the universal masses for the scalar SUSY fields and gauginos. In addition, we do not assume any fixed energy scale for the $U(1)^{\prime}$ symmetry breaking, thus the VEVs of the MSSM singlets are also included in the free parameter set of this class of models. We will consider only the solutions which are consistent with the several constraints such as the mass bounds, constraints from rare $B-$meson decays, Planck measurements on the relic abundance of LSP neutralino within $5\sigma$ and the exclusion bounds from the direct detection experiments. After confronting such solutions with several analyses, we proceed to discuss possible signal processes suitable with the light Higgs bosons through their associated production with $Z$ and photons. {In our work, on the other hand, we briefly discuss the associated production of light Higgs bosons with $Z$, while we consider those involving photons in details. The rest of the paper is organized as follows: We first briefly describe the model in Section \ref{sec:model} in terms of the possible charge assignments, the Higgs bosons and their mixing in \ref{subsec:hbosons}, and their possible signals in Section \ref{subsec:monoX}. After summarizing our scanning procedure the employed experimental constraints in Section \ref{sec:scan}, we will compare the light Higgs boson solutions with several analyses in Section \ref{subsec:SMdecays} and discuss the possible prospects to test/probe such solutions through their associated productions with photons in Section \ref{subsec:monophoton}.} Finally we conclude and summarize our results in Section \ref{sec:conc}.

\section{Model Description}
\label{sec:model}

In this section we will briefly describe the models in the class of Secluded UMSSM with an emphasize on its salient features which are relevant to our work. The general gauge group of UMSSM can be written as $SU(3)_{C}\times SU(2)_{L}\times U(1)_{Y}\times U(1)^{\prime}$. Such models can arise from the symmetry breaking in $E_{6}$ and/or $SO(10)$ models, as mentioned in Eq.(\ref{eq:breaking}), but its character does not, in general, have to follow this symmetry breaking chain. Indeed, the anomaly cancellation can be satisfied with different charge assignments. Table \ref{tab:benchcharges} exemplifies some sets of possible charge assignments which are also of interest from phenomenology point of view \cite{Hicyilmaz:2021khy}. 

\begin{table}[h!]
\caption{Sets of $U(1)^{\prime}$ charges which satisfy the conditions of anomaly cancellation and gauge invariance in the Secluded $U(1)^{\prime}$ model.}
\centering
\setstretch{2.0}
\scalebox{0.82}{
\begin{tabular}{|c|c|c|c|c|c|c|c|}
\hline
$U(1)^{\prime}$ Charges             & Set 1 & Set 2 & Set 3 & Set 4 & Set 5& Set 6 & Set 7 \\ \hline
                        $Q_{Q}^{\prime}$ &0.0    &-0.05  &-0.1   &0.2     &0.15 &0.1     &0.1\\
                        $Q_{U}^{\prime}$ &-0.45&0.45    &-0.05   &-0.35 &0.15 &-0.45 &0.1\\
                        $Q_{D}^{\prime}$ &0.9    &-0.8      &0.7     &-0.5   &0.0 &0.55 &0.0\\                                                             
                        $Q_{L}^{\prime}$ &0.15   &0.0    &0.45   &-0.75 &-0.3 &-0.2 &-0.2\\
                        $Q_{N}^{\prime}$ &-0.6  &0.4     &-0.6   &0.6    &0.6 &-0.15 &0.4\\
                        $Q_{E}^{\prime}$ &0.75  &-0.85     &0.15 &0.45   &0.45 &0.85 &0.3\\
                        $Q_{H_u}^{\prime}$ &0.45&-0.4   &0.15 &0.15   &-0.3 &0.35 &-0.2\\
                        $Q_{H_d}^{\prime}$ &-0.9&0.85  &-0.6   &0.3    &-0.15 &-0.65 &-0.1\\
                        $Q_{S}^{\prime}$ &0.45  &-0.45     &0.45   &-0.45 &0.45 &0.3 &0.3\\
                        $Q_{S_1}^{\prime}$ &0.45&-0.45    &0.45   &-0.45  &0.45&0.3 &0.3\\
                        $Q_{S_2}^{\prime}$ &0.45&-0.45   &0.45    &-0.45  &0.45&0.3 &0.3\\
                        $Q_{S_3}^{\prime}$ &-0.9&0.9     &-0.9     &0.9    &-0.9&-0.6 &-0.6\\
                        \hline
\end{tabular}}
\label{tab:benchcharges}
\end{table}
These charge sets cannot be obtained through the mixing of $U(1)_{\chi}$ and $U(1)_{\psi}$ thorough $\theta_{E_{6}}$. Note that the exotics are not included in Table \ref{tab:benchcharges}, but we consider the minimal choice of the exotics including three color triplets with $Y_{\mathcal{Q}}=-1/3$ and two color singlets with $Y_{\mathcal{L}}=-1$, where $Y_{\mathcal{Q}}$ and $Y_{\mathcal{L}}$ stand for their hypercharges. Since the MSSM fields are non-trivially charged under $U(1)^{\prime}$, the UMSSM sector significantly interfere with the low scale phenomena and considerably alter the implications. One of the advantages of the presence of the secluded sector is involving consistently heavy $Z^{\prime}$ without shifting the masses in this sector to the multi-TeV scales.

Even though the spectrum involves a heavy $Z^{\prime}$, it can still interfere in the low scale phenomenology through a possible gauge kinetic mixing between $U(1)_{Y}$ and $U(1)^{\prime}$. In the case of the gauge mixing, the covariant derivative has a non-canonical form, which enhance the mixing between the MSSM Higgs fields and the singlets at tree-level. However, the current analyses have severely bounded the gauge kinetic mixing from above as $\xi \lesssim 3\times 10^{-4}$ from searches over different decay modes of $Z^{\prime}$ \cite{Pankov:2019yzr,Bobovnikov:2018fwt,CMS:2016wev}. In addition, the DM direct detection experiments can provide an upper bound through the photon and $Z-$boson abundance in the DM scattering processes \cite{Lao:2020inc}. Considering these strict bounds, we do assume the gauge kinetic mixing to be zero in our work, since it does negligibly affect the low scale implications.

\subsection{The Higgs bosons in Secluded UMSSM}
\label{subsec:hbosons}

The presence of the MSSM singlet fields ($S$ and $S_{i}$, where $i=1,2,3$) significantly extends the MSSM Higgs sector in a way that the physical Higgs spectrum include six CP-even and four CP-odd Higgs bosons. The electroweak symmetry breaking in these models is realized in a similar way, but since the Higgs potential, in general, involves mixing terms between the MSSM Higgs doublets and singlets, the electroweak symmetry breaking is now connected to the $U(1)^{\prime}$ symmetry breaking. The general Higgs potential can be written as $V_{tree}=V_F+V_D+V_{soft}$, where  \cite{Chiang:2008ud}:

\begin{equation}
\setstretch{2.5}
\begin{array}{ll}
V_{F} & =|\lambda|^{2}\left[|H_{d}H_{u}|^{2} +|S|^{2}(H_{d}^{\dagger}H_{d}+H_{u}^{\dagger}H_{u})\right]+\dfrac{|\kappa|^{2}}{9}\left(|S_{1}S_{2}|^{2}+|S_{2}S_{3}|^{2}+|S_{1}S_{3}|^{2} \right]~, \\
V_{D} = & \dfrac{g_{1}^{2}+g_{2}^{2}}{8}\left(H_{d}^{\dagger}H_{d}-H_{u}^{\dagger}H_{u} \right)^2+ \dfrac{g_{2}^{2}}{2}|H_{d}^{\dagger}H_{u}|^{2} \\
 & +\dfrac{g'^{2}_{1}}{2}\left(Q'_{H_{d}}H_{d}^{\dagger}H_{d}+Q'_{H_{u}}H_{u}^{\dagger}H_{u}+Q'_{S}|S|^{2}+\displaystyle \sum_{i=1}^{3}Q'_{S_{i}}|S_{i}|^{2}
 \right)^{2}~, \\ 
 V_{soft}= & m_{H_{d}}^{2}H_{d}^{\dagger}H_{d}
+ m_{H_{u}}^{2}H_{u}^{\dagger}H_{u}+m_{S}^{2}|S|^{2}+\displaystyle \sum_{i=1}^{3}m_{S_{i}}^{2}|S_{i}|^{2} \\
 & -\left( \lambda A_{\lambda} SH_{d}H_{u} + \dfrac{\kappa}{3}A_{\kappa}S_{1}S_{2}S_{3}+h.c.\right)~ \\ 
 & -\left( m_{SS_{1}}^{2}SS_{1}+m_{SS_{2}}^{2}SS_{2}+m_{S_{1}S_{2}}^{2}S^{\dagger}_{1}S_{2}+h.c.\right)~,
\end{array}
\end{equation}

where $g_{1}$, $g_{2}$ and $g_{1}^{\prime}$ are the gauge couplings of $SU(2)_{L}$, $U(1)_{Y}$ and $U(1)^{\prime}$, respectively. In the Higgs potential generated by the $F-$terms ($V_{F}$), the secluded singlets do not mix with the others, while the $U(1)^{\prime}$ sector interferes in MSSM through the tree-level coupling between $S$ and the MSSM Higgs fields, Moreover, the $D-$term generated potential ($V_{D}$) mixes all the Higgs fields through their $U(1)^{\prime}$ charges. The SSB terms given in $V_{soft}$ rather mix $S$ with the MSSM Higgs fields in one side, and with the secluded singlets in the other side. In this context, the mixing among the Higgs fields are not trivial, and even singlet scalar bosons can contribute to the physical observables at the low scale. The SSB terms also include mixing terms among the singlets as given in the last line of $V_{soft}$. These terms also necessary to break unwanted global $U(1)$ symmetries which require $Q^{\prime}_{S_{1}}=Q^{\prime}_{S_{2}}=-Q^{\prime}_{S}$. 


After the symmetry breaking, the non-zero elements of the mass-square matrix for the CP-even Higgs bosons can be obtained as follows;

\begin{equation*}
\setstretch{2}
\Scale[0.75]{
\begin{array}{l}
M_{11}^{2} = \dfrac{(g_{1}^{2}+g_{2}^{2})v_{d}^{2}}{4}+g'^{2}_{1}Q_{H_{d}}^{\prime2}v_{d}^{2}+\dfrac{A_{\lambda}\lambda v_{S}v_{u}}{\sqrt{2}v_{d}}~, \\
M_{12}^{2}=-\dfrac{A_{\lambda}\lambda v_{S}}{\sqrt{2}}- \dfrac{(g_{1}^{2}+g_{2}^{2})v_{d}v_{u}}{4}+\left(\lambda^{2} + g'^{2}_{1}Q'_{H_{d}}Q'_{H_{u}} \right)v_{d}v_{u}~, \\
M_{13}^{2}=\lambda^{2}v_{d}v_{s}+g'^{2}_{1}Q'_{H_{d}}Q'_{S}v_{d}v_{S}-\dfrac{A_{\lambda}\lambda v_{u}}{\sqrt{2}}~, \\
M_{1i+3}^{2}=g'^{2}_{1}Q'_{H_{d}}Q'_{S_{i}}v_{d}v_{S_{i}}~; \hspace{0.3in} i=1,2,3~, \\
M_{22}^{2}=\dfrac{A_{\lambda}\lambda v_{d}v_{S}}{\sqrt{2}v_{u}}+\dfrac{1}{4}\left(g_{1}^{2}+g_{2}^{2} + 4g'^{2}_{1}Q'_{H_{u}}\right) v_{u}^{2}~, \\
M_{23}^{2}=-\dfrac{A_{\lambda}\lambda v_{d}}{\sqrt{2}} + \left(\lambda^{2}+g'^{2}_{1}Q'_{H_{u}}Q'_{S}\right)v_{u}v_{S}~,\\
M_{2i+3}^{2}=g'^{2}_{1}Q'_{H_{u}}Q'_{S_{i}}v_{u}v_{S_{i}}~; \hspace{0.3in} i=1,2,3~,
\end{array}\hspace{0.5cm}
\begin{array}{l}
M_{33}^{2} = \dfrac{1}{2v_{S}}\left(2g'^{2}_{1}Q_{S}^{\prime2}v_{s}^{3} -2m_{SS_{1}}^{2}v_{S_{1}}- 2m_{SS_{2}}^{2}v_{S_{2}}+\sqrt{2}A_{\lambda}\lambda v_{d}v_{u}\right)~, \\
M_{3i+3}^{2}=m_{SS_{i}}^{2}+g'^{2}_{1}Q'_{S}Q'_{S_{i}}v_{S}v_{S_{i}}~\hspace{0.3in} i=1,2,3~{\rm and}~m_{SS_{3}}=0~, \\
M_{44}^{2} = \dfrac{1}{2v_{S_{1}}}\left(2g'^{2}_{1}Q_{S_{1}}^{\prime2}v_{S_{1}}^{3}-2m_{SS_{1}}^{2}v_{S}+\sqrt{2}A_{\kappa}\kappa v_{S_{2}}v_{S_{3}} \right)~, \\
M_{45}^{2} = \dfrac{1}{9}\kappa^{2}v_{S_{1}}v_{S_{2}}+g'^{2}_{1}Q'_{S_{1}}Q'_{S_{2}}v_{S_{1}}v_{S_{2}}-\dfrac{A_{\kappa}\kappa v_{S_{3}}}{\sqrt{2}}~, \\
M_{46}^{2} = \dfrac{1}{9}\left(\kappa^{2}+9g'^{2}_{1}Q'_{S_{1}}Q'_{S_{3}} \right) v_{S_{1}}v_{S_{3}}   - \dfrac{A_{\kappa}\kappa v_{S_{2}}}{\sqrt{2}}~, \\
M_{55}^{2}=\dfrac{1}{2v_{S_{2}}}\left(2g'^{2}_{1}Q_{S_{2}}^{\prime2}v_{S_{2}}^{3}-2m_{SS_{2}}^{2}v_{S}+\sqrt{2}A_{\kappa}\kappa v_{S_{1}}v_{S_{3}}\right)~, \\
M_{56}^{2}=\dfrac{1}{9}\left( \kappa^{2}+9g'^{2}_{1}Q'_{S_{2}}Q'_{S_{3}}\right) v_{S_{2}}v_{S_{3}}-\dfrac{A_{\kappa}\kappa v_{S_{1}}}{\sqrt{2}}~, \\
M_{66}^{2} = g'^{2}_{1}Q_{S_{3}}^{\prime2}v_{S_{3}}^{2} + \dfrac{A_{\kappa}\kappa
v_{S_{1}}v_{S_{2}}}{\sqrt{2}v_{S_{3}}}~.
\end{array}}
\end{equation*}

The first effect can be noted in the SM-like Higgs boson. The tree-level SM-like Higgs boson mass is enhanced by the gauge coupling and charges associated with the $U(1)^{\prime}$ group as well as the tree-level coupling $\lambda$. In contrast to the tree-level limit on the SM-like Higgs boson in MSSM, UMSSM can accommodate a consistent Higgs boson mass even at the tree-level. Even though it receives the loop contributions mainly from the stop sector in UMSSM \cite{Langacker:1998tc,Carena:1995wu,Demir:1998dk}, the typical spectra do not have to involve heavy stops and/or large trilinear couplings \cite{Hicyilmaz:2017nzo}. 

Similarly, the non-zero elements in the mass-square matrix of the CP-odd Higgs bosons are

\begin{equation*}
\setstretch{2}
\begin{array}{l}
P_{11}^{2} = \dfrac{A_{\lambda}\lambda v_{S}v_{u}}{\sqrt{2}v_{d}}~,~ P_{12}^{2} = \dfrac{A_{\lambda}\lambda v_{S}}{\sqrt{2}}~,~ P_{13}^{2} =
\dfrac{A_{\lambda}\lambda v_{u}}{\sqrt{2}}~, \\
P^{2}_{22} = \dfrac{A_{\lambda}\lambda v_{d}v_{S}}{\sqrt{2}v_{u}}~, P^{2}_{23}\dfrac{A_{\lambda}\lambda v_{d}}{\sqrt{2}}~, \\
P^{2}_{33} = \dfrac{1}{2v_{S}}\left(-2m_{SS_{1}}^{2}v_{S_{1}}-2m_{SS_{2}}^{2}v_{S_{2}}+\sqrt{2}A_{\lambda}\lambda v_{d}v_{u}  \right)~,~P^{2}_{34}=-m^{2}_{SS_{1}}~,~P^{2}_{35}=-m^{2}_{SS_{2}}~, \\
P^{2}_{44}=\dfrac{1}{2v_{S_{1}}} \left(-2m_{SS_{1}}^{2}v_{S}+\sqrt{2}A_{\kappa}\kappa v_{S_{2}}v_{S_{3}} \right)~,~ p^{2}_{45}=\dfrac{A_{\kappa}\kappa v_{S_{3}}}{\sqrt{2}}~,~P^{2}_{46}=\dfrac{A_{\kappa}\kappa v_{S_{2}}}{\sqrt{2}} \\
P^{2}_{55} = \dfrac{1}{2v_{S_{2}}}\left(-2m_{SS_{2}}^{2}v_{S} + \sqrt{2}A_{\kappa}\kappa v_{S_{1}}v_{S_{3}}  \right)~,~P^{2}_{56}=\dfrac{A_{\kappa}\kappa v_{S_{1}}}{\sqrt{2}}~,~P^{2}_{66}=\dfrac{A_{\kappa}\kappa v_{S_{1}}v_{S_{2}}}{\sqrt{2}v_{S_{3}}}~.
\end{array}
\end{equation*}

The tree-level masses of the singlet Higgs bosons are proportional to $m_{SS_{1}}^{2}$ and $m_{SS_{2}}^{2}$ if $v_{S}\simeq v_{S_{i}}$, while it is suppressed more if $v_{S_{3}} \gg v_{S},v_{S_{1,2}}$ which typically happens to involve sufficiently heavy $Z^{\prime}$ boson in the physical spectrum. It has been shown that the Secluded UMSSM models can accommodate two relatively lighter CP-odd Higgs bosons which are lighter than about 250 GeV and 700 GeV, respectively. The DM relic density constraint reduces these bounds further as $m_{A_{1}}\lesssim 225$ GeV and $m_{A_{2}}\lesssim 300$ GeV  \cite{Hicyilmaz:2021khy}. Even though these scalars are mostly singlets, their presence in the spectrum can mimic through their mixing with the MSSM Higgs fields. The problem with the light scalar states can arise from the rare $B-$meson decays such as $B_{s}\rightarrow \mu^{+}\mu^{-}$ and $B_{s}\rightarrow X_{s}\gamma$. Since, the experimental measurements of these rare $B-$meson decays are in a strong agreement with the SM predictions \cite{HFLAV:2022pwe,CMS:2020rox,Belle-II:2022hys}, these rare processes of $B-$meson provide strong constraints on the predictions of the model. In addition, these singlets can be probed with the analyses over the Higs boson decays into these scalars \cite{CMS:2018nsh,CMS:2018qvj,CMS:2019idx}.

These light singlets also take important part in the DM implications, since a light singlet LSP, as a candidate of DM, needs to undergo right amount of the self-annihilation processes to yield a relic density compatible with the latest measurements of Planck satellite within $5\sigma$. The Higgs bosons participate in such annihilations of LSP, and reduce its relic density significantly when they happen to be resonances with the LSP neutralino. When the lightest CP-even Higgs boson is required to be the SM-like Higgs boson, it can play important role when $m_{\tilde{\chi}_{1}^{0}} \simeq 60$ GeV. The width of these processes is controlled by the coupling between $S$ and the MSSM Higgs fields ($\lambda$), and it can be constrained severely by the latest measurements on the invisible decays of the SM-like Higgs bosons (${\rm BR}(h\rightarrow {\rm invisible}) \lesssim 10\%$ \cite{CMS:2019bke,Aaboud:2018sfi,Sirunyan:2018owy,Aad:2015pla,Chatrchyan:2008aa,Khachatryan:2016whc}). Therefore, realizing the correct relic abundance of LSP needs also light Higgs bosons resonances. 

In addition to the light CP-odd Higgs bosons, as mentioned above, the models in the Secluded UMSSM class involve six CP-even Higgs bosons. Leaving the MSSM CP-even Higgs bosons out there remain four mostly singlet CP-even Higgs bosons in the spectrum. These CP-even Higgs boson can weigh up to about 2 TeV in the parameter space of Secluded UMSSM (described in Section \ref{sec:scan}), they can interfere in the processes with the final states of MSSM particles through their mixing with the MSSM Higgs fields. These singlet scalars can receive some exclusive impacts from the recent collider analyses. These analyses can also provide some potential to probe such Higgs bosons, since their sensitivity has been constantly being improved.

\subsection{Mono-X Signals involving Singlet Higgs Bosons}
\label{subsec:monoX}

\begin{figure}[h!]
\centering
\includegraphics[scale=0.6]{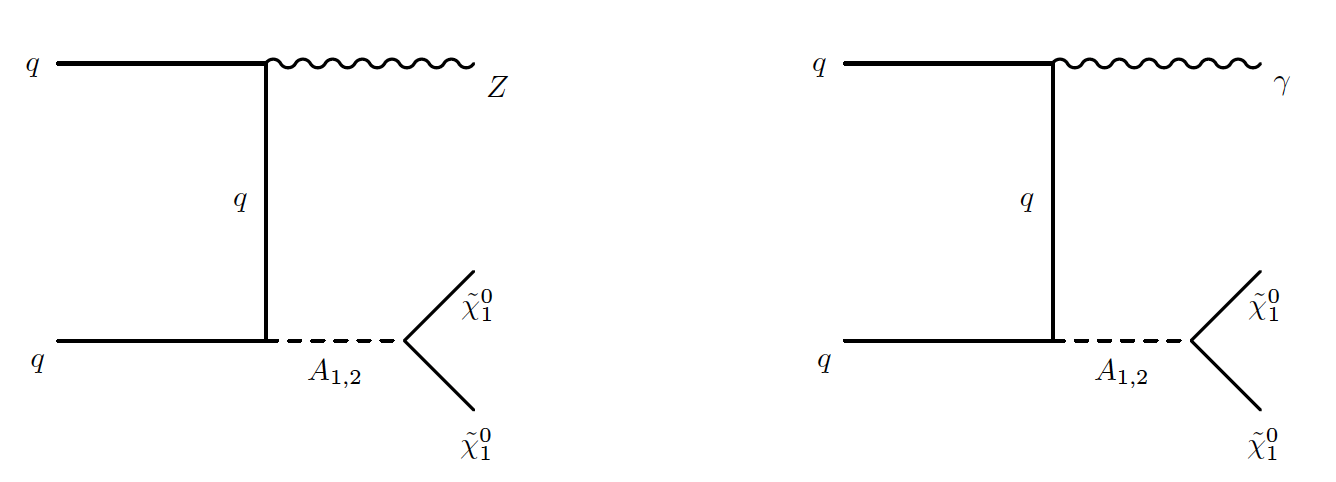}
\caption{Possible signal processes involving the singlet scalars ($A_{1,2}$) associated with the $Z-$boson (left) and photon (right).}
\label{fig:signals}
\end{figure}

Even though all the constraints mentioned above can still allow the light scalar states (see, for instance, \cite{Cici:2019zir,Moretti:2022okg,Arhrib:2021xmc,Abdelalim:2020xfk,Ahriche:2022aoj,Li:2022etb,Khalil:2021afa}) when they are almost purely singlet, there is still a cracked window for a small but non-zero mixing with the MSSM Higgs fields. Despite the challenges in proposing possible signal processes with significant cross-sections, the sensitivity of the current collider experiments and associated analyses can track possible signatures of such light singlets. These analyses can be extended to include the associated productions of these Higgs bosons with some SM particles. Indeed, these associated productions can be one of the main ingredient, if the singlet scalars decay into the LSP, which can be traced only through the missing energy in the colliders, while the involved SM particles can form some visible final states. Such processes can be summarized with the associated production of the $Z-$boson and photon as shown in Figure \ref{fig:signals}. The left diagram represents the signal processes in which one of the initial quarks radiates a $Z-$boson before producing the singlet CP-odd Higgs boson, which decays into a pair of LSP neutralinos subsequently. In the region of our interest, the LSP is also formed by the MSSM singlet neutralinos, and the decay branching ratio for $A\rightarrow \tilde{\chi}_{1}^{0}\tilde{\chi}_{1}^{0}$ is almost 100\%. The second diagram represents similar signal processes in which the $Z-$boson is replaced by a photon.

If one follows the signal processes involving $Z-$boson as shown in the left diagram of Figure \ref{fig:signals}, the main SM background is formed by $ZZ$ and $WW$ productions. One of the $Z-$bosons in the $ZZ$ production decays into a pair of neutrinos forming the missing energy, while the other decays into leptons or quarks. In the $WW$ production, the $W-$bosons decay into a charged lepton and its neutrino. However, due to the possibility of poor reconstructions of jets, $Z/W + $jets can also contribute to the background processes. Similarly, if the energetic electrons (or muons) in the final states escape from the detection, they also contribute to the missing energy, thus, $ZZ\rightarrow llll$ also becomes a relevant process in the background analysis as well as the $WZ$ background \cite{ATLAS:2021gcn}. Considering the QCD uncertainties, $Z+$jets form a significant part of the background. The uncertainties in mono-Z signal analyses increase further if $Z-$boson decays into a pair of $\tau-$leptons.

On the other hand, $Z+$jets processes may lose their significance if one considers the other signal processes involving mono-photon as shown in the right diagram of Figure \ref{fig:signals}. The main SM background is formed mostly by the processes involving $Z-$ or $W-$ bosons produced associated with a photon. In these processes $Z-$ boson decays into a pair of neutrinos forming the missing transverse energy ($\cancel{E}_{T}$), while $W$ follows decays into a charged lepton and its neutrino. Due to the possibility of misidentification of electrons and/or jets, also $\gamma + $jets and the processes in which a $Z-$boson decays into a pair of leptons can also contribute to the background processes \cite{ATLAS:2018jsf}. However, applying appropriate criteria can reduce such signals. For instance, requiring $\cancel{E}_{T}/(\sqrt{\textstyle{\small \sum}E_{T})} \geq 8.5$ GeV$^{1/2}$ reduces the $\gamma +$jets events less than $10\%$ of the background processes \cite{ATLAS:2017nga}. Similarly, misidentifying electrons (and muons) as fake photons and/or their poor reconstructions lead to $ZZ\rightarrow llll~(ll\nu\nu)$, $Z\gamma \rightarrow ll\gamma$ and $WW\rightarrow ll\nu\nu$  events contributing to the background as well, where $l$ denotes the electron or muon. However, comparing with the observed number of events, such processes can form a small portion of the SM background, and they can be accounted for subleading processes. Considering these advantages in the case of mono-photon final states, we will consider the mono-photon signal processes in probing the light scalar states in our work.

Note that there are also other potential signal processes involving mono-Higgs and its decay products in the final state. These processes arise from either a direct product of a MSSM CP-odd Higgs boson or an off-shell $Z-$boson production decaying into a scalar singlet together with the SM-like Higgs boson. However, due to the severe constraints from rare $B-$meson decays on the CP-odd Higgs boson, its production is rather suppressed by the heavy masses of CP-odd Higgs boson. Similarly, those involving off-shell $Z-$boson can be significant if the models yields a large gauge kinetic mixing and relatively lighter $Z^{\prime}$. In this context, it is also suppressed in our model since we assume zero gauge kinetic mixing and $Z^{\prime}$ is set to be heavy.



\section{Scanning Procedure and Experimental Constraints}
\label{sec:scan}

We perform scans in the parameter space of the Secluded UMSSM models built by imposing the universal boundary conditions at $\mgut$. The free parameters and their ranges are given in Table \ref{tab:scan_lim}, where $m_{0}$ is the soft supersymmetry breaking (SSB) mass term assigned to all the scalars including the Higgs fields, and $M_{1/2}$ stands for SSB mass term for all the gauginos. The VEVs of the MSSM Higgs fields are parametrised by  $\tan\beta \equiv v_{u}/v_{d}$. $\lambda$ measures the interactions between the scalar singlet field $S$ and the MSSM Higgs fields as defined earlier, while $\kappa$ is the self-interaction coupling in the secluded sector. The universal trilinear coupling between the MSSM matter fields and the Higgs fields is varied in terms of its ratio to the universal SSB mass term of the scalar fields in order to avoid the color and/or charge breaking minima of the scalar potential \cite{Ellwanger:1999bv,Camargo-Molina:2013sta}. On the other hand, we directly vary the trilinear coupling associated with $\lambda$ and $\kappa$, and we follow the usual convention for the interaction strength as $T_{\lambda}\equiv A_{\lambda}\lambda$ and $T_{\kappa} = A_{\kappa}\kappa$. Finally $h_{\nu}$ stands for the Yukawa coupling between $H_{u}$ and the right-handed neutrino fields. Considering the tiny neutrino masses established by the experiments \cite{Super-Kamiokande:2010orq}, if one assumes TeV scale right-handed neutrinos, $h_{\nu}$ is restricted to be at the order of $10^{-7}$ or smaller \cite{Abbas:2007ag}.

\begin{table}[h]
        \setlength\tabcolsep{20pt}
        \renewcommand{\arraystretch}{2.0}
        \begin{tabular}{|c|c||c|c|}
                \hline
                Parameter      & Scanned range& Parameter      & Scanned range \\
                \hline
                $m_0$        & $[0, 10]$ TeV     & $v_S$   & $[1, 20]$ TeV  \\
                $M_{1/2}$   & $[0, 10]$ TeV     & $v_{S_1}$   & $[3, 20]$ TeV  \\
                $\tan\beta$   & $[1, 50]$        & $v_{S_2}$  & $[3, 20$] TeV  \\
                $A_0/m0$   & $[-3, 3]$   & $v_{S_3}$  & $[3, 20]$ TeV \\
                $\lambda$   & $[0.01, 0.5]$ & $A_\lambda$   & $[0, 10]$ TeV  \\
                $\kappa$   & $[0.1, 1.5]$ & $A_\kappa$   & $[-10, 0]$ TeV   \\
                $h_\nu $   & $[10^{-11}, 10^{-7}]$  &    &  \\
                \hline
        \end{tabular}
        $\vspace{0.5cm}$
        \caption{\label{tab:scan_lim} The set of the free parameters of the Secluded UMSSM and their ranges.}
\end{table}

In our scans, we employ SPheno 4.0.4 package \cite{Porod:2003um,Porod:2011nf,Goodsell:2014bna} generated with SARAH 4.14.3 \cite{Staub:2013tta,Staub:2015kfa,Goodsell:2014bna}. In this package, the unification scale of the gauge couplings is calculated by running the renormalization group equations (RGEs) from $M_{Z}$ to the higher scales by inputting the weak scale values of the MSSM gauge and Yukawa couplings. The unification scale is determined as a high scale at which the unification condition ($g_{3}\approx g_{1}=g_{2}=g_{1}^{\prime}$) is realized, where $g_{3}$, $g_{2}$ and $g_{1}$ are the MSSM gauge couplings for $SU(3)_{C}$, $SU(2)_{L}$ and $U(1)_{Y}$ respectively, and $g_{1}^{\prime}$ denotes the gauge coupling associated with the $U(1)^{\prime}$ gauge group. After the unification scale is calculated, all the SSB parameters are inputted in the RGEs and they are run from $\mgut$ back to $M_{Z}$ together with the gauge and Yukawa couplings. 

In analysing the generated data, we impose several constraints such as the LEP2 bounds on the masses \cite{Patrignani:2016xqp} and the mass bound on the gluino mass from the current analyses \cite{ATLAS:2019fag}. We also require the solutions to yield the lightest CP-even Higgs boson to be the SM-like Higgs boson with consistent mass and decay modes reported by ATLAS \cite{ATLAS:2012yve,ATLAS:2017bxr,ATLAS:2019aqa,ATLAS:2018uoi} and CMS \cite{CMS:2012qbp,CMS:2017dib,CMS:2015hra,CMS:2013zmy} collaborations.
{Note that the Higgs boson mass constraint yield a strong impact especially on the stop mass as it requires $m_{\tilde{t}_{1}} \gtrsim 1$ TeV, when the gluino weighs around 8 TeV, while it excludes the solutions with $m_{\tilde{t}_{1}} \lesssim 3$ TeV when gluino mass is realized at around 2.1 TeV \cite{Hicyilmaz:2021khy}. Such an impact implies also much heavier squarks of the first two families, which yield solutions consistent with the squark-gluino searches \cite{ATLAS:2020syg,ATLAS:2017xvf}}. In addition to the mass bounds and Higgs decays, we apply the constraints from rare $B-$meson decays such as $ {\rm BR}(B \rightarrow X_{s} \gamma) $ \cite{Amhis:2012bh}, $ {\rm BR}(B_s \rightarrow \mu^+ \mu^-) $ \cite{Aaij:2012nna} and $ {\rm BR}(B_u\rightarrow\tau \nu_{\tau}) $ \cite{Asner:2010qj}.


\begin{table}
\caption{The experimental constraints employed in our analyses.}
\centering
\setstretch{2.0}
\begin{tabular}{|c|c|c|}
\hline
Observable & Constraint & Ref. \\ \hline
$m_{\tilde{\chi}_{1}^{\pm}},m_{\tilde{\tau}}$ & $\geq 100$ GeV & \cite{Patrignani:2016xqp} \\
$m_{\tilde{g}}$ & $\geq 2.1$ TeV & \cite{ATLAS:2019fag} \\
$m_{h}$ & $[122-128]$ GeV & \cite{ATLAS:2012yve,CMS:2013btf} \\
$M_{Z^{\prime}}$ & $\geq 4$ TeV & \cite{Pankov:2019yzr,Bobovnikov:2018fwt,CMS:2016wev,Lao:2020inc,ATLAS:2019erb,CMS:2019tbu} \\
${\rm BR}(B\rightarrow X_{s}\gamma)$ & $[2.99 - 3.87]\times 10^{-4}~(2\sigma)$ & \cite{Amhis:2012bh} \\
${\rm BR}(B_{s}\rightarrow \mu^{+}\mu^{-})$ & $[0.8 - 6.2]\times 10^{-9}~(2\sigma)$ & \cite{Aaij:2012nna} \\
$\dfrac{{\rm BR}(B_{u}\rightarrow \tau \nu_{\tau})_{{\rm Secluded~}U(1)^{\prime}}}{{\rm BR}(B_{u}\rightarrow \tau \nu_{\tau})_{{\rm SM}}}$ & $[0.15 - 2.41]~(2\sigma)$ & \cite{Asner:2010qj} \\
$\Omega_{{\rm CDM}}h^{2}$ & $[0.114 - 0.126]~(5\sigma)$ & \cite{Aghanim:2018eyx}\\ \hline
\end{tabular}
\label{tab:constraints}
\end{table}

Note that we have employed about 3 GeV uncertainty in the Higgs boson mass calculation. Despite its significantly precise experimental measurement, most of the uncertainty comes from its theoretical calculation due to the large uncertainties in the top-quark mass, strong coupling, the mixing in the stop sector, which yields an overall uncertainty in the Higgs boson mass up to about 3 GeV \cite{Degrassi:2002fi,Allanach:2004rh,Athron:2016fuq,Drechsel:2016htw,Bagnaschi:2017xid,Bahl:2019hmm}. In addition to its mass, we also require that its invisible decays cannot exceed $11\%$ \cite{ATLAS:2023tkt}. After requiring the consistent mass within the interval given in Table \ref{tab:constraints} and consistent decay modes, the composition of the lightest CP-even Higgs boson is observed as $|ZH_{11}|^{2} +|ZH_{12}|^{2} \gtrsim 80\%$, where $Z_{H}$ is the matrix which diagonalises the mass-square matrix of the CP-even Higgs bosons, the first subindex stands for the lightest CP-even Higgs boson, while the second subindices 1 and 2 correspond to $H_{d}$ and $H_{u}$ respectively.

{After applying all these constraints, we accept only the solutions which are consistent with all of them including the DM relic density and the bounds from the direct dark matter detection experiments \cite{LUX:2016ggv,XENON:2018voc,PandaX-II:2017hlx}. In applying the direct detection constraint, we take the results from the XENON1T experiment \cite{XENON:2018voc}. Even though the current exclusion depends on the mass of LSP, in the mass scales of LSP realized in our scans the current bound mostly requires the solutions to yield spin-independent scattering cross-section of DM less than about $10^{-10}$ pb (detailed discussion including other LSP species can be found in \cite{Hicyilmaz:2021khy}).}

\section{Constraining and Probing the Light Higgs Bosons}

As has been discussed above, the singlet scalars arising in the secluded sector are allowed to mix significantly with the MSSM Higgs fields. Despite the set of severe constraints applied in or analyses as summarized in Section \ref{sec:scan}, there is still a considerable possibility for these light Higgs bosons to mimic in some collider analyses. These processes can involve them decaying into the SM vector bosons or charged leptons such as $\tau$, whose visible final states have been analyzed with a considerable precision. In this section, we will first discuss the allowed decay modes of the light Higgs bosons into the SM particles and discuss the impact from the recent analyses as well as the possibility to probe such light scalars through these analyses. In the second part, we will consider the production of these light Higgs bosons associated with a photon and discuss the current sensitivity of the collider experiments and the possible projections to probe such processes in the future colliders.

\subsection{Decay Modes involving the SM Particles}
\label{subsec:SMdecays}

\begin{figure}[t!]\hspace{-0.6cm}
\subfigure{\includegraphics[scale=0.5]{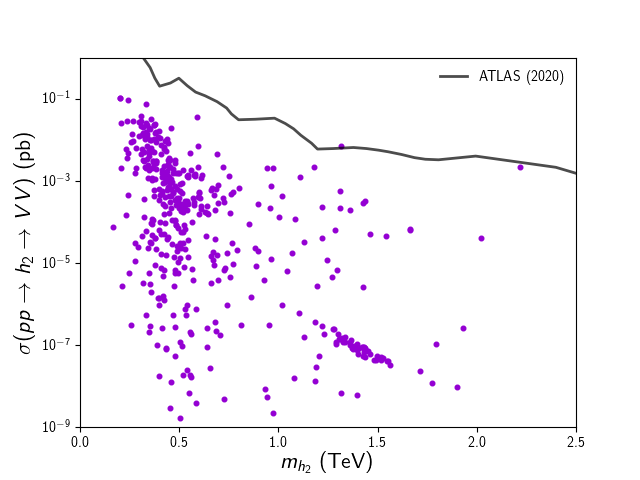}}%
\subfigure{\includegraphics[scale=0.5]{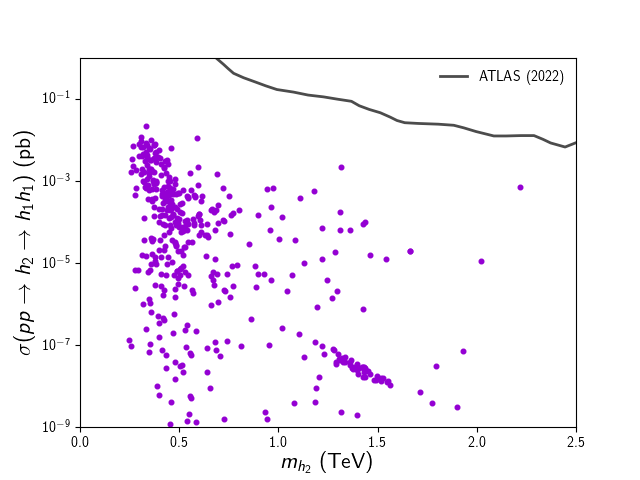}}
\caption{The cross-section of the heavy CP-even Higgs bosons decaying into a pair of vector bosons $V=W^{\pm},Z$ (left), and a pair of the SM-like Higgs bosons (right). All the points are consistent with the experimental constraints summarized in Section \ref{sec:scan}. The solid black curves represent the current bounds on these processes from the experimental analyses \cite{ATLAS:2020fry,ATLAS:2022hwc}.}
\label{fig:CPevenHtoSM}
\end{figure}

The coupling of the singlet Higgs bosons to the SM particles are suppressed by the small mixing with the MSSM Higgs fields in their composition, and it affects their decay modes into the SM particles as well as their productions in the collider experiments. However, despite its smallness, the sensitivity of the current analyses are able to track them even if they slightly contribute to the processes with the final states formed by the SM particles. On the other hand such processes are not convenient to probe the light CP-odd Higgs bosons. Even though these light scalars can develop considerable couplings with the SM particles, the soft transverse momenta ($p_{T} \lesssim 25$ GeV) suppress the visibility of the signal over the background processes.

We first show in Figure \ref{fig:CPevenHtoSM} the cross-sections of heavy CP-even Higgs boson decays into a pair of vector bosons $V=W^{\pm},Z$ (left), and a pair of the SM-like Higgs bosons (right). All the points are consistent with the experimental constraints summarized in Section \ref{sec:scan}. The solid black curves represent the current bounds on these processes from the experimental analyses \cite{ATLAS:2020fry,ATLAS:2022hwc}. Our results show that $m_{h_{2}}$ can lie from about 200 GeV to 2.3 TeV. In this mass interval, a large mixing among the singlet scalars and the MSSM Higgs fields can be consistent with the experimental constraints, and such a large mixing can receive some impact from the analyses over the processes involving a pair of SM gauge bosons. As shown in the right panel of Figure \ref{fig:CPevenHtoSM}, the current analyses are capable to exclude such solutions even when $m_{h_{2}} \gtrsim 1.3$ TeV. The implications of the Secluded UMSSM for $h_{2}$, when it decays into a pair of the SM gauge bosons, are placed slightly lower than but close by the current exclusion curve, and one can expect these solutions to be probed in the upcoming experimental analyses. There are also solutions in the same mass interval, which yield very small cross-sections, these solutions are characterized with a large percentage of the MSSM singlet scalar fields in the composition of $h_{2}$. A similar discussion can be followed for $h_{2}$ in its decay processes into a pair of the SM-like Higgs bosons. Despite the similar sensitivities of the experimental analyses over these processes, the Secluded UMSSM predicts rather smaller cross-sections for the $h_{2}\rightarrow h_{1}h_{1}$, but some solutions can be probed by the experimental analyses in near future, as they require some further improvements in the experimental sensitivity. The small cross-section solutions correspond to the solutions in which $h_{2}$ is mostly formed by the singlet scalars with small $\lambda$. 

\begin{figure}[t!]\hspace{-0.6cm}
\subfigure{\includegraphics[scale=0.5]{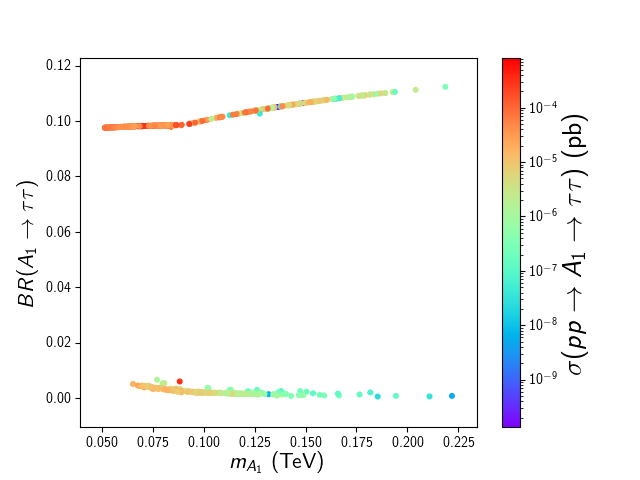}}%
\subfigure{\includegraphics[scale=0.5]{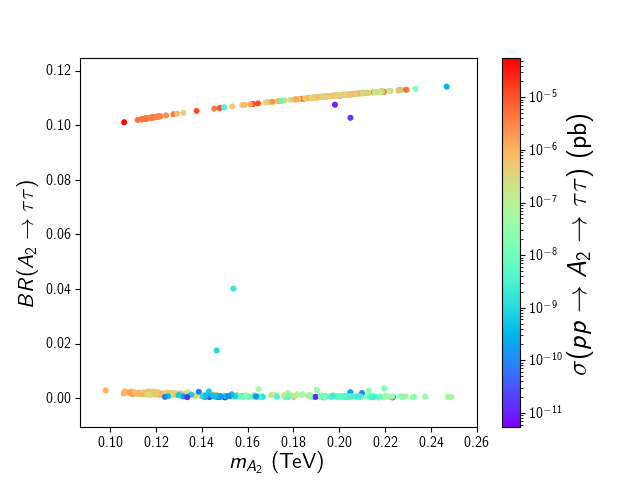}}
\caption{Branching fractions for $\tau\tau$ decays of the light CP-odd Higgs bosons in correlation with their masses for $A_{1}$ (left) and $A_{2}$ (right). All the points are consistent with the experimental analyses employed in our analyses. The solutions are colored with respect to the total cross-section of $pp\rightarrow A_{i}\rightarrow \tau\tau$, $i=1,2$, which is given with a color bar in the side for each panel.}
\label{fig:Atautau}
\end{figure}

We continue our discussion on constraining and/or probing the Higgs sector of the Secluded UMSSM with the current experimental analyses with its decay mode to a pair of $\tau-$leptons. Figure \ref{fig:Atautau} displays the branching fractions of the light CP-odd Higgs bosons for the $A_{i}\rightarrow \tau\tau$ decays in correlation with the masses of these light scalars. All the points are consistent with the experimental analyses employed in our analyses. The solutions are colored with respect to the total cross-section of $pp\rightarrow A_{i}\rightarrow \tau\tau$, $i=1,2$, which is given with a color bar in the side for each panel. According to our results, the lightest CP-odd Higgs bosons can decay into a pair of $\tau-$leptons at the rate of $10-12\%$. Considering ${\rm BR}(A_{1}\rightarrow \tau\tau)$ together with the production cross-section of $A_{1}$, the total cross-section of $pp\rightarrow A_{1}\rightarrow \tau\tau$ can be realized as high as about $10^{-4}$ pb when $m_{A_{1}} \lesssim 100$ GeV. The total cross-section becomes negligible when $A_{1}$ is heavier than 100 GeV. Similar results can be observed also the second lightest CP-odd Higgs boson as shown in the ${\rm BR}(A_{2}\rightarrow \tau\tau)-m_{A_{2}}$ plane of Figure \ref{fig:Atautau}. $A_{2}$ happens to be heavier than about 100 GeV in our scans, and the total cross-section for $pp\rightarrow A_{2}\rightarrow \tau\tau$ can be realized at about $10^{-6}$ pb when its mass is about 210 GeV. The relevant analyses have been reported by the ATLAS collaboration for these decays \cite{ATLAS:2020zms}, and they rather probe the scalars decaying into a $\tau-$pair when their masses are around 200 GeV or heavier. The total cross-section of the observed $\tau\tau$ events are at the order of $10^{-2}$pb, which is way above our results for $pp\rightarrow A_{i}\rightarrow \tau\tau$ events in the Secluded UMSSM models.

{Even though we observe two different branching ratios for $A_{i}\rightarrow \tau\tau$ decay mode in the results represented in Figure 3, these two branches can be distinguished by the decays of the light CP-odd Higgs bosons into a pair of LSP neutralinos. When $A_{i}\rightarrow \tilde{\chi}_{1}^{0} \tilde{\chi}_{1}^{0}$ mode is not kinematically allowed (i.e. $m_{A_{i}} < 2m_{\tilde{\chi}_{1}^{0}}$), these Higgs bosons mostly decay into the SM particles, and the branching ratios are observed at the rate of about 10\% for their $\tau$ decays (the branch in Figure 3 with ${\rm BR}(A_{i}\rightarrow \tau\tau) \simeq 10-12\%$), and about $90\%$ when the final states are formed by $b-$quarks. On the other hand, when $A_{i}$ mass is large enough to allow the decays into LSP neutralinos, $A_{i}\rightarrow \tilde{\chi}_{1}^{0} \tilde{\chi}_{1}^{0}$ happens at a rate about $99\%$. Even though the both cases yield different rate for the $A_{i}\rightarrow {\rm SM ~ SM}$ decays, the widths of these decays in both cases are realized at the order about $10^{-7}-10^{-8}$ GeV.}

\begin{figure}[t!]\hspace{-0.6cm}
\subfigure{\includegraphics[scale=0.5]{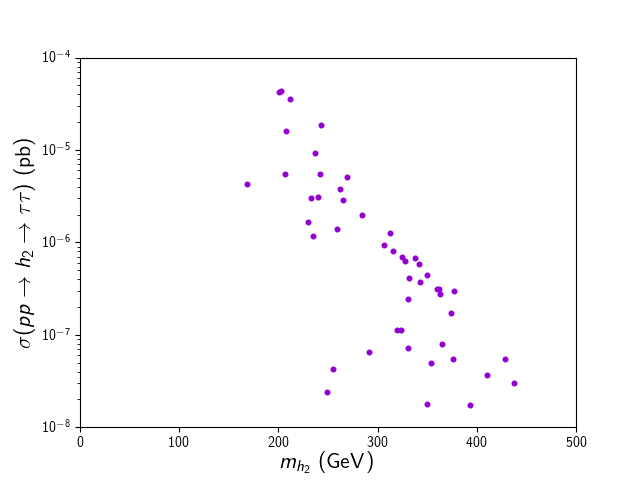}}%
\subfigure{\includegraphics[scale=0.5]{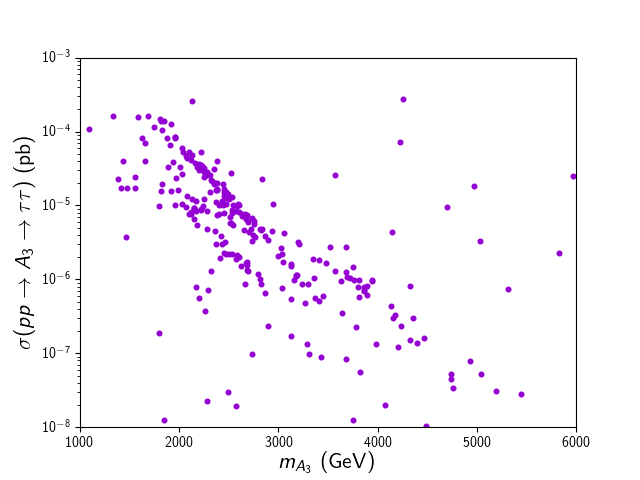}}
\caption{$\tau\tau$ events involving the relatively heavier Higgs bosons $h_{2}$ (left) and $A_{3}$ (right) in correlation with their masses. All the points are consistent with the experimental constraints applied in our analyses.}
\label{fig:HHtautau}
\end{figure}

Similar discussion can be followed also for the relatively heavier Higgs bosons and their $\tau\tau$ decays which are shown in Figure \ref{fig:HHtautau}. The larger cross-section for these events in our scans is realized as about $10^{-4}$ pb for the heavy CP-even Higgs boson as shown in the left panel. {Since the MSSM Higgs fields cannot mix with the singlet fields in forming the lighter Higgs bosons, they mostly interfere in composition of the heavier Higgs states. Their significant mixing in these heavier Higgs bosons enhances their $\tau\tau$ decays, while it also yields a stronger impact from the constraints from rare $B-$meson decays ($B_{s}\rightarrow \mu \mu$ and $B\rightarrow X_{s}\gamma$). Due to these constraints, $A_{3}$ cannot be lighter than about 1 TeV, while the $A_{3}\rightarrow \tau\tau$ events can yield a cross-section as high as about $3\times 10^{-4}$ pb, as shown in the right plane of Figure \ref{fig:HHtautau}.}


\subsection{The associated production of Light Scalars Associated with Photon}
\label{subsec:monophoton}

As we discussed in the previous subsection, the events in which the scalars decay into a pair of the SM particles can be applicable and/or suitable when these scalars have masses about 200 GeV or heavier. Besides, these analyses require the solutions to have large mixing between the singlet and MSSM Higgs fields, while they lose their impacts when the mixing is small. In the cases of small mixing, the singlet scalars can decay into a pair of LSP with a large rate, and when LSP is mostly formed by the fermionic singlets of Secluded UMSSM, this decay rate can be observed at about $100\%$. However, these decays manifest itself through the missing energies in the collider experiments. The small mixing also suppresses the production of these light scalars in collisions, and thus the processes leading to totally missing energy in the final states are severely suppressed by the SM background. Even though the SM background remains strong, there might be other processes, which can leave traceable marks in the analyses. One of the typical topology for such events involves associated production of the light singlet scalars with a Standard Model particle, in which the singlet scalar decays totally into a pair of LSP neutralinos, while the involved SM particle forms a visible final state through its decays.

In this context, we consider the possible signal processes involving these light singlet scalars through their production associated with a photon, which is shown with the right diagram in Figure \ref{fig:signals}. The SM background relevant to these events is mainly formed by the $Z-$boson pair production processes in which one $Z-$boson decays into a pair for neutrinos forming the missing energy, while the other decays into leptons or quarks leading to visible final states. However, misidentification and/or poor reconstraction of the leptons and jets allow other processes such as $Z\rightarrow ll\gamma$, $W\rightarrow ll\nu\nu$ and $\gamma + $jets processes to contribute to the SM background. Among these, some selection rules on the missing and total transverse energy can reduce the events involving $\gamma +$jets significantly. Moreover, the rest of the processes can contribute only at the subleading level in comparison with the main $ZZ$ background. 

{Before proceeding in discussion of our results for the mono-photon events mediated by singlet scalars, one needs to consider how the signals can manifest itself in the experiments. As shown in previous section, our model can yield two scalars yielding a photon in the final state with a missing energy. Distinguishing these two scalars from each other depends on the sensitivity of to the missing energy and/or mass re-constraction of the particles involved in the decays. Even though the efficiency is very low (around $10\%$), the detectors are sensitive when an event has $\cancel{E}_{T} \gtrsim 50$ GeV. As the efficiency increases with the missing energy, $\cancel{E}_{T}$ can be reconstructed with about $60\%$ efficiency when it is about 100 GeV, and the efficiency reaches to about $100\%$ when $\cancel{E}_{T} \gtrsim 150$ GeV \cite{CMS:2016ljj}. In addition, the missing mass calculator (MMC) method \cite{Elagin:2010aw} can be employed which is using the missing energy and the observables of the visible states in the final state of the decays to distinguish the masses of the particles involving in the decay cascade under concern. Using this technique, one can distinguish two states of different masses eventually decaying in the process. This technique can distinguish two particles if their mass difference is about 15 GeV or more when the particles decay into $\tau-$leptons together with neutrinos \cite{Elagin:2010aw}. The method can be applied in our case by considering the missing energy, the photon $p_{T}$ and the angle between these two parameters. However, considering the efficiency of missing energy re-constraction mentioned above, the sensitivity to the masses of the particles can be worsened for cases of the light Higgs bosons. In this context, we assume only one of the singlet CP-odd Higgs bosons can dominantly contribute to the mono-photon events, while the other's contributions can be minor or negligible.}

\begin{figure}[t!] \hspace{-0.6cm}
\subfigure{\includegraphics[scale=0.5]{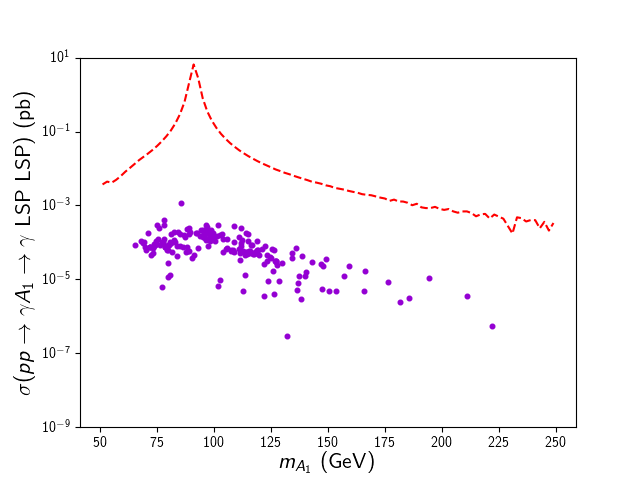}}%
\subfigure{\includegraphics[scale=0.5]{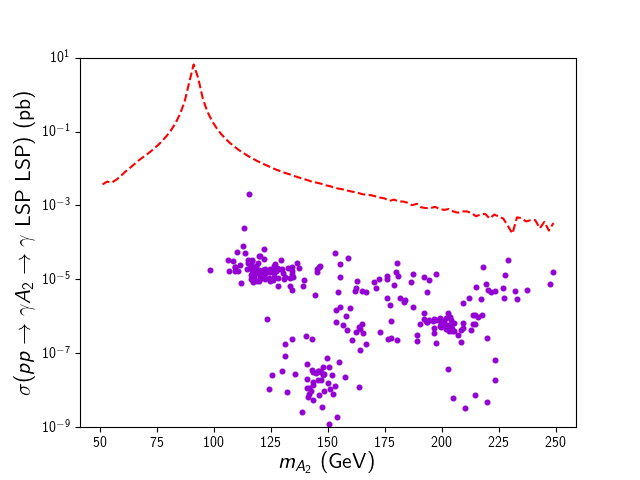}}
\caption{The mono-photon signal processes involving $A_{1}$ (left) and $A_{2}$ (right) in correlation with their masses. All points are consistent with the experimental constraints applied in our analyses. The dashed red curve represents the total cross-section of the relevant SM background processes except those with $\gamma +$jets with respect to the missing transverse energy where we assume $\cancel{E}_{T} \approx m_{A_{i}}$, $i=1,2$.}
\label{fig:sigmonophoton}
\end{figure}

We show our results in Figure \ref{fig:sigmonophoton} for the cross-section of the mono-photon signal {at 14 TeV center of mass energy} involving one of these light singlet CP-odd Higgs bosons in correlation with their masses. All points are consistent with the experimental constraints applied in our analyses. The dashed red curve represents the total cross-section of the relevant SM background processes except those with $\gamma +$jets with respect to the missing transverse energy where we assume $\cancel{E}_{T} \approx m_{A_{i}}$, $i=1,2$. We employ the following approximation in calculation of the total signal cross-section:

\begin{equation}
\sigma(pp\rightarrow \gamma A_{i}\rightarrow \gamma~ {\rm LSP~ LSP})\simeq \sigma(pp\rightarrow \gamma A_{i})\times {\rm BR}(A_{i}\rightarrow {\rm LSP~ LSP})~,
\end{equation}
where the pair of LSP manifests itself as the missing transverse energy. {The associated production of $A_{i}$ with the photon is calculated by $ MadGraph5 $ \cite{Alwall:2014hca}.} Note that this approximation yields about $1\%$ or less uncertainty in our calculations compared to the full matrix element consideration \cite{Altin:2020qmu}. In these events, the missing transverse energy is approximately determined by the mass of the CP-odd Higgs boson involved in the process. The left panel of Figure \ref{fig:sigmonophoton} shows the signal cross-section for those involving $A_{1}$, and the largest cross-section can be realized at the order of $10^{-3}$ pb when $m_{A_{1}} \simeq 100 $ GeV. With the missing energy as $\cancel{E}_{T}\simeq 100$ GeV, the SM background processes also yield a peak at which the cross-section of the total background is about 10 pb. However, if one applies a cut on the missing transverse energy as $\cancel{E}_{T}\gtrsim 130$ GeV, the background processes can be significantly reduced to about $10^{-3}$ pb. Similar reduction in the background can be realized also with a cut as $\cancel{E}_{T}\lesssim 80$ GeV. Similarly, one can realize the mono-photon signal processes involving $A_{1}$ with a cross-section as large as about $10^{-3}$ pb; however, such solutions can be observed with some missing transverse energies for which the background processes can be sufficiently suppressed. {In sum, we observe the possible largest cross-sections at about $1\times 10^{-3}$ pb when $m_{A_{1}}\simeq 85$ GeV for the mono-photon events involving $A_{1}$ (left plane of {Figure \ref{fig:sigmonophoton}}), and at about $2\times 10^{-3}$ pb for those involving $A_{2}$ when $m_{A_{2}} \simeq 115$ GeV (right panel of {Figure \ref{fig:sigmonophoton}}).}

\begin{figure}[t!]\hspace{-0.6cm}
\subfigure{\includegraphics[scale=0.5]{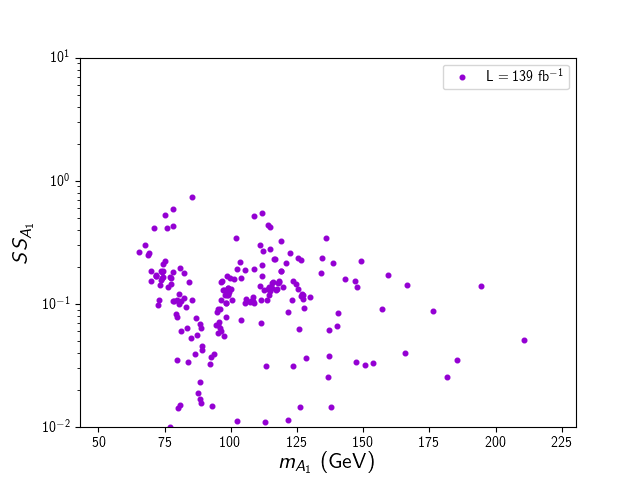}}%
\subfigure{\includegraphics[scale=0.5]{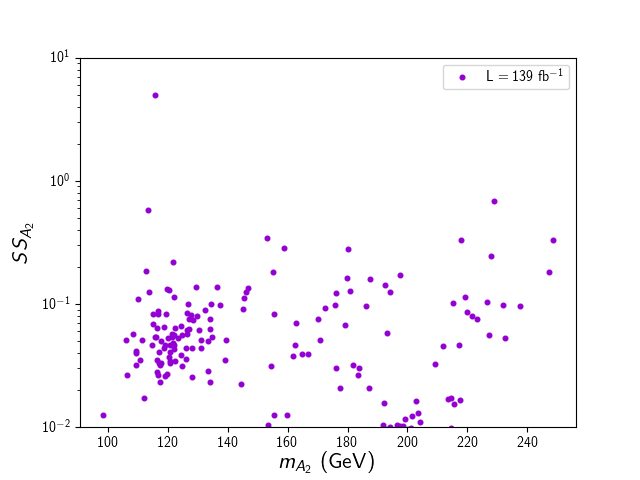}}
\caption{The significance for the signal processes involving $A_{1}$ (left) and $A_{2}$ (right). All the points are consistent with the experimental constraints applied in our analyses. The significance is calculated approximately by $SS_{A_{i}}=S_{A_{i}}/\sqrt{S_{A_{i}} + B}$, where $S_{A_{i}}$ and $B$ represent the number of events for the signal and background processes, respectively.}
\label{fig:monosign}
\end{figure}

Even though the signal processes in the class of Secluded UMSSM models yield cross-sections lower than the background processes, they can still considerably contribute to the total mono-photon events. In this context, the significance of the signal processes can be a better measure to explore if they can yield visible traces or under which circumstances they can lead to visible events. We display our results for the signal significance for the signal processes involving $A_{1}$ (left) and $A_{2}$ (right) in Figure \ref{fig:monosign}. All the points are consistent with the experimental constraints applied in our analyses. The significance of the signal processes involving $A_{i}$ ($SS_{A_{i}}$) is calculated approximately by $SS_{A_{i}}=S_{A_{i}}/\sqrt{S_{A_{i}} + B}$, where $S_{A_{i}}$ and $B$ represent the number of events for the signal and background processes, respectively. The largest significance for the mono-photon events involving $A_{1}$ is realized to be about 0.8 when $m_{A_{1}} \simeq 80$ GeV. Even though the solutions with a significance less than 1 can escape from the detection in the current experiments, one can expect a probe for them at around 68\% Confidence Level (CL) at the end of Run-3 experiments, whose targeted integral luminosity is 400 fb$^{-1}$. The sensitivity of their probe will more likely reach to or exceed the 95\% CL when the LHC operates with higher luminosity (at 3000 fb$^{-1}$ of High-Luminosity LHC (HL-LHC), for instance). Although the heavier $A_{1}$ solutions yield relatively lower significance, they are still considerable, and one can expect that the near future collider experiments including those in Run-3 phase of LHC will be able to probe $A_{1}$ solutions up to about $m_{A_{1}} \simeq 200$ GeV.

Furthermore, the signal processes involving $A_{2}$ can yield even greater significance, since they can lead to large cross-section out of the mass interval where the background processes yield a peak. The $SS_{A_{2}}-m_{A_{2}}$ plot shows that there exist solutions with $m_{A_{2}} \simeq 120$ GeV that can show themselved with a large significance ($SS_{A_{2}} \sim 6$). Such solutions can be considered to be excluded already. However, depending on their mixing with the MSSM Higgs fields, one can also realize solutions whose significance is slightly lower than 1, but they are still considerable. Similarly these solutions can yield visible events during the period of Run-3 operation of LHC, and the near future experiments are expected to strengthen the impact to probe such solutions up to $m_{A_{2}} \simeq 250$ GeV. 

Note that the mass scales commented in our discussions are obtained by the constraints summarized in Section \ref{sec:scan}. These upper bounds for the light singlet scalars have rather obtained through the consideration of the DM observables. If one does not assume that the LSP neutralino is accounted for the DM observables, the mass scales can be extended beyond our bounds, and the near future experiments are able to probe a larger parameter space of the Secluded UMSSM models than that commented in our work.

\begin{table}[h!]
	\caption{The benchmark points for mono-photon signals involving $A_{1}$. The points are selected to be consistent with the experimental constraints. All masses and trilinear couplings are given in TeV, the decay widths in GeV and the cross-sections in $ pb$.}
	\centering
	\setstretch{2.0}
	\scalebox{0.7}{
		\begin{tabular}{|c|c|c|c|c|}
			\hline
			Parameters &  BM1 & BM2 & BM3 & BM4 \\ \hline
			$\tan\beta$ &22.6&15.7&11.2&32.2\\
			$v_{s}$ &6.25&5.81&3.91&5.57\\
			$( \lambda,\ \kappa)$&(0.054,\ 0.39)&(0.056,\ 1.19)&(0.094,\ 1.44)&(0.11,\ 0.86)\\
			$(A_{\lambda},\ A_{\kappa})$ &(1.1,\ -3.8)&(1.5,\ -8.9)&(1.6,\ -9.3)&(8.1,\ -2.8)\\
			$(v_{s_1},\ v_{s_2},\ v_{s_3}) $ &(9.47,\ 13.6,\ 12.1)&(6.3,\ 10.8,\ 14.5)&(8.57,\ 11.4,\ 10.9)&(5.61,\ 4.55,\ 6.26)\\
			$(m_{0})$ &3.053&7.477&9.417&4.565\\
			$(M_{1/2})$ &3.762&3.986&2.714&7.576\\
			$(A_{0})$ &-1.404&2.322&4.251&0.331\\
			Charge Set  & Set 2 & Set 3 & Set 3 &Set 5\\ \hline
			$(m_{H_1},\ m_{H_2})$ & 
			(0.1237,\,0.424) &
			(0.1244,\,0.367) &
			(0.1233,\,0.570) &
			(0.1273,\,0.543)\\
			$(m_{A_1},\ m_{A_2})$ & 
			(0.085,\,0.124) &
			(0.210,\,0.223) &
			(0.136,\,0.193) &
			(0.194,\,0.223)\\
			$(m_{\tilde\chi^0_1},\ m_{\tilde\chi^0_2})$ & 
			(0.041,\,0.246) &
			(0.111,\,0.198) &
			(0.060,\,0.267) &
			(0.094,\,0.462\\
			$m_{H^{\pm}}$ &2.503&2.402&2.181&10.763\\ \hline
			$\sigma(pp\rightarrow h_{2}\rightarrow VV) $&$ 5.94\times 10^{-3}$&$ 3.08\times 10^{-3}$&$ 3.74 \times 10^{-4}$&$ 	3.60 \times 10^{-3}$\\
			$\sigma(pp\rightarrow h_{2}\rightarrow h_{1}h_{1}) $&$ 2.08 \times 10^{-3}$&$ 1.01 \times 10^{-3}$&$ 1.15 \times 10^{-4}$&$ 2.70 \times 10^{-5}$\\
			$\sigma(pp\rightarrow \gamma A_{1}\rightarrow \gamma~ {\rm LSP~ LSP})$ &$ 1.16 \times 10^{-3} $&$ 2.50 \times 10^{-4}$&$ 7.20 \times 10^{-5}$&$ 1.10 \times 10^{-5}$\\
			$\sigma(pp\rightarrow \gamma A_{2}\rightarrow \gamma~ {\rm LSP~ LSP})$ &$ 1.58 \times 10^{-5} $&$ 7.84 \times 10^{-7}$&$ 4.54 \times 10^{-6}$&$ 6.44 \times 10^{-8}$\\
			$\Gamma_{A_{1}}$ &$ 4.94 \times 10^{-5} $&$ 1.90 \times 10^{-4}$&$ 9.79 \times 10^{-4}$&$ 7.70 \times 10^{-4}$\\
			$\Gamma_{A_{2}}$  &$ 1.03 \times 10^{-4} $&$ 6.03 \times 10^{-5}$&$ 9.09 \times 10^{-4}$&$ 1.40 \times 10^{-5}$\\ \hline
	\end{tabular}}
	\label{tab:benchmark1}
\end{table}

\begin{figure}[ht!]\hspace{-0.6cm}
\subfigure{\includegraphics[scale=0.5]{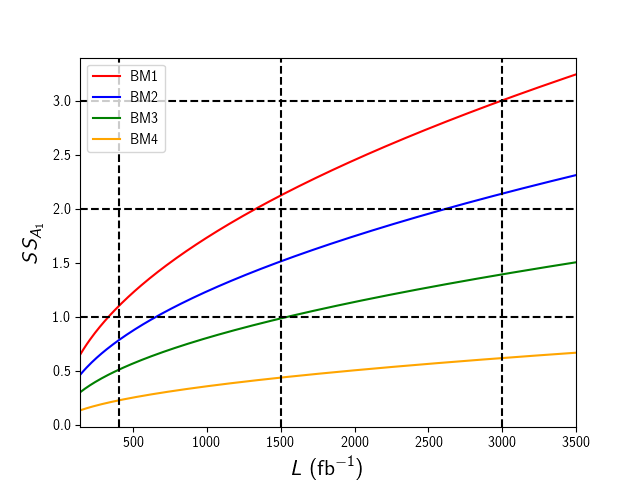}}%
\subfigure{\includegraphics[scale=0.5]{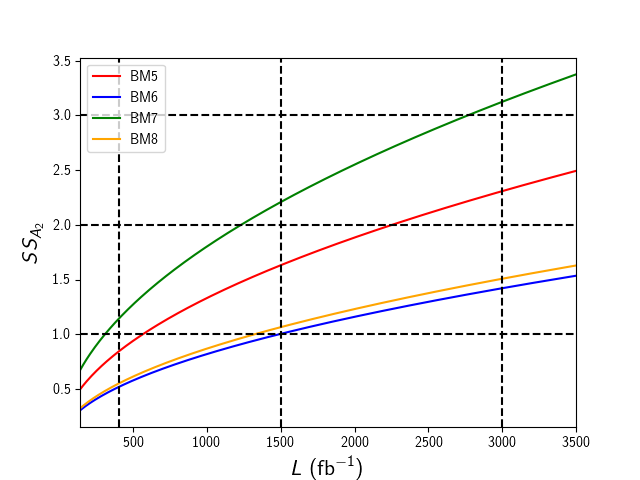}}
\caption{Significance evolution of the benchmark points given in Table \ref{tab:benchmark1} (left) and Table \ref{tab:benchmark2} (right). Each curve represents a benchmark point whose color coding is given in the legend. The horizontal dashed lines show the significance values of 1, 2 and 3 from bottom to top, while the vertical dashed lines represent the targeted integral luminosity values at the end of phases of LHC for Run-3 (400 fb$^{-1}$), Run-4 (1500 fb $^{-1}$) and HL-LHC (3000 fb$^{-1}$) from left to right in both panels.}
\label{fig:signevol}
\end{figure}

To summarize our discussion about the significance of the signal processes, we display eight benchmark points (BMs). Table \ref{tab:benchmark1} display four solutions for the signal processes involving $A_{1}$, while those involving $A_{2}$ is given in Table \ref{tab:benchmark2}. These points are selected to be consistent with the experimental constraints applied in our analyses including the CP-even Higgs boson decays discussed in Section \ref{subsec:SMdecays}. All the masses and dimensionful trilinear couplings are given in TeV, while the cross-sections are represented in pb. We also display the significance evolutions for $SS_{A_{1}}$ (left) and $SS_{A_{2}}$ (right) in Figure \ref{fig:signevol}. Each curve represents a benchmark point whose color coding is given in the legend. The horizontal dashed lines show the significance values of 1, 2 and 3 from bottom to top, while the vertical dashed lines represent the targeted integral luminosity values at the end of phases of LHC for Run-3 (400 fb$^{-1}$), Run-4 (1500 fb $^{-1}$) and HL-LHC (3000 fb$^{-1}$) from left to right in both panels.  

\begin{table}[h!]
	\caption{The benchmark points for mono-photon signals involving $A_{2}$. The points are selected to be consistent with the experimental constraints. All masses are given in TeV, the decay widths in GeV and the cross-sections in $pb $.}
	\centering
	\setstretch{2.0}
	\scalebox{0.7}{
		\begin{tabular}{|c|c|c|c|c|}
			\hline
			Parameters &  BM5 & BM6 & BM7 & BM8 \\ \hline
			$\tan\beta$ &18.7&24.4&44.2&13.3\\
			$v_{s}$ &6.55&6.34&5.05&3.15\\
			$( \lambda,\ \kappa)$&(0.057,\ 0.42)&(0.1,\ 0.96)&(0.077,\ 1.26)&(0.17,\ 1.29)\\
			$(A_{\lambda},\ A_{\kappa})$ &(0.7,\ -4.7)&(3.1,\ -4.8)&(0.5,\ -9.9)&(7.6,\ -4.7)\\
			$(v_{s_1},\ v_{s_2},\ v_{s_3}) $ &(9.2,\ 13.1,\ 14.2)&(10.9,\ 9.9,\ 14.4)&(10.1,\ 18.8,\ 12.4)&(6.68,\ 6.97,\ 5.01)\\
			$(m_{0})$ &5.465&4.815&5.430&7.414\\
			$(M_{1/2})$ &4.402&7.643&7.942&7.550\\
			$(A_{0})$ &-4.040&-2.683&13.046&18.988\\
			Charge Set  & Set 7 & Set 4 & Set 6 &Set 1\\ \hline
			$(m_{H_1},\ m_{H_2})$ & 
			(0.1263,\,0.364) &
			(0.1235,\,0.378) &
			(0.1229,\,0.515) &
			(0.1278,\,0.505)\\
			$(m_{A_1},\ m_{A_2})$ & 
			(0.071,\,0.113) &
			(0.136,\,0.153) &
			(0.101,\,0.228) &
			(0.185,\,0.248)\\
			$(m_{\tilde\chi^0_1},\ m_{\tilde\chi^0_2})$ & 
			(0.034,\,0.276) &
			(0.067,\,0.501) &
			(0.050,\,0.288) &
			(0.059,\,0.402)\\
			$m_{H^{\pm}}$ &1.911&6.011&2.133&6.294\\ \hline
			$\sigma(pp\rightarrow h_{2}\rightarrow VV) $&$ 8.16 \times 10^{-5}$&$ 2.41 \times 10^{-2}$&$ 9.81 \times 10^{-8}$&$ 1.21 \times 10^{-7}	$\\
			$\sigma(pp\rightarrow h_{2}\rightarrow h_{1}h_{1}) $&$ 3.31\times 10^{-5}$&$ 7.35 \times 10^{-3}$&$ 3.32 \times 10^{-8}$&$ 3.03 \times 10^{-8}	$\\
			$\sigma(pp\rightarrow \gamma A_{1}\rightarrow \gamma~ {\rm LSP~ LSP})$ &$ 7.57 \times 10^{-5} $&$ 5.23 \times 10^{-6}$&$ 6.41\times 10^{-6}$&$ 3.18 \times 10^{-6}	$\\
			$\sigma(pp\rightarrow \gamma A_{2}\rightarrow \gamma~ {\rm LSP~ LSP})$ &$ 2.50 \times 10^{-4} $&$ 5.03 \times 10^{-5}$&$ 3.27\times 10^{-5}$&$1.63\times 10^{-5} 	$\\
			$\Gamma_{A_{1}}$ &$ 1.95 \times 10^{-5} $&$ 3.59 \times 10^{-5}$&$ 5.83 \times 10^{-6}$&$ 2.47 \times 10^{-4}$\\
			$\Gamma_{A_{2}}$  &$ 9.81 \times 10^{-5} $&$ 4.90 \times 10^{-4}$&$ 1.49 \times 10^{-3}$&$ 5.29 \times 10^{-3}$\\ \hline
	\end{tabular}}
	\label{tab:benchmark2}
\end{table}

Table \ref{tab:benchmark1} displays the solutions which can be traced through the mono-photon signals involving $A_{1}$ in $85 \lesssim m_{A_{1}} \lesssim 200$ GeV. The largest signal cross-section is observed at the order of $10^{-3}$ pb when $m_{A_{1}} \simeq 85$ GeV, while it reduces to the magnitudes of order $10^{-4}-10^{-5}$ pb with increasing $A_{1}$ mass. All these points yield $SS_{A_{1}}$ less than 1 in the current experiments. However, BM1 and BM2 have $SS_{A_{1}}$ greater than 0.5, and they are expected to be probed soon before the Run-3 experiments of LHC come to an end. Even though BM3 yields relatively lower significance, the near future experiments will potentially be able to probe such solutions, as well. The required luminosity values to probe these solutions can be seen from the left panel of Figure \ref{fig:signevol}, and they are approximately 300, 500 and 1200 fb$^{-1}$ for BM1, BM2 and BM3 respectively. On the other hand, one can read from the curves that the solutions of the signal cross-sections less than about $5\times 10^{-5}$ pb may need further upgrades in the collider experiments to be probed, since their significance remains less than 1 even in the HL-LHC experiments as exemplified by BM4 whose significance evolution is shown with the orange curve in the left panel of Figure \ref{fig:signevol}. Note that these benchmark points do not exemplify the solutions with $90\lesssim m_{A_{1}} \lesssim 100$ GeV, since the number of the background processes in this interval yield a peak and strongly suppress the signals.

A similar discussion can be followed for the mono-photon signals with $A_{2}$, which is exemplified with the benchmark points given in Table \ref{tab:benchmark2}. Even though these processes yield lower cross-sections in comparison with those involving $A_{1}$, since $A_{2}$ has heavier masses ($120 \lesssim m_{A_{2}} \lesssim 230$ GeV), the signal processes exhibit a stronger potential to be distinguished from the background processes in the missing transverse energy. Thus, these benchmark points yield a comparable significance with those given in Table \ref{tab:benchmark1}. As shown in the right panel of Figure \ref{fig:signevol}, BM5 and BM7 (red and green curves, respectively) are expected to be probed at the end of Run-3. On the other hand BM7 and BM8 yield signal cross-sections less than $3\times 10^{-5}$ pb. However, the lower bound on the signal cross-sections in these mass scales is approximately $0.5 \times 10^{-5}$ pb, and these solutions are expected to be available to be analyzed in the collider experiments when the integrated luminosity is about 1400 fb$^{-1}$ or greater.


\begin{figure}[h!]
\subfigure{\includegraphics[scale=0.45]{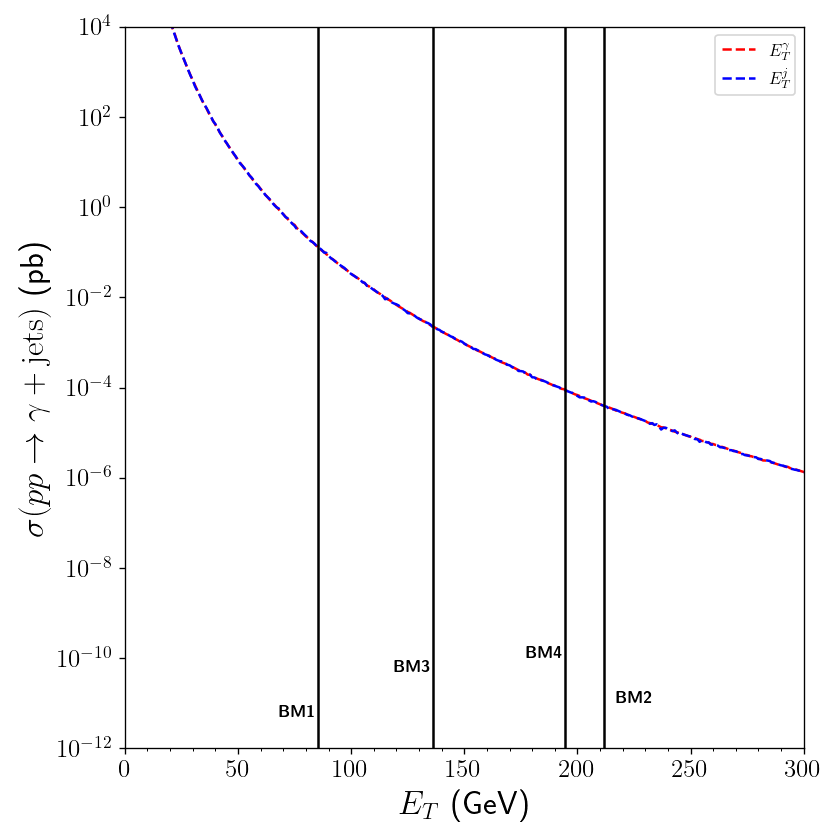}}%
\subfigure{\includegraphics[scale=0.45]{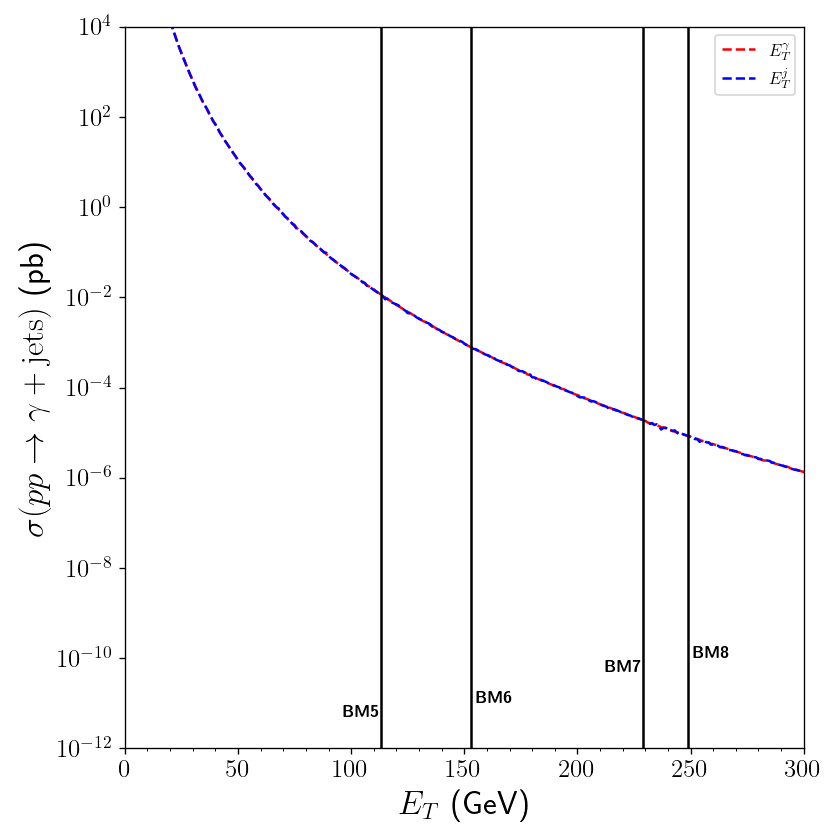}}
\caption{The cross-section of the $\gamma + $jets processes in correlation with the transverse energy in the final state. The blue dashed curve displays the transverse energy of the jets, while the $E_{T}^{\gamma}$ is represented by the red dashed curve. Since it overlaps with the jet transverse energy it is barely visible. The vertical solid lines show the approximate missing transverse energy of the signal processes for each benchmark point given in Table 4 and Table 5.}
\label{fig:crossET}
\end{figure}

{Note that even though we left discussions on the $\gamma +$jets processes to the end, the significance curves involve these processes as well. Although they might not be considered as a main constitute of the background processes, they are also of a special interest \cite{ATLAS:2022raw,ATLAS:2023yrt,Petersson:2012dp,Orimoto:2015ueg,ATLAS:2017nga,CMS:2018ffd}. Suppressing the $\gamma +$jets processes without altering the signal cross-section is not a straightforward process, since the photon behaves similarly in both the background and signal processes. Both classes of the processes yield large cross-sections when the photon transverse momentum is low ($p_{T}^{\gamma} \simeq 20$ GeV). However in these cases the missing transverse energy can serve as an observable which distinguishes the signal from the background processes. Assuming the jets to be nearly massless, the transverse energy of the jets can be observed approximately equal to $p_{T}^{\gamma}$, while since the signal processes involve massive particles (LSP neutralinos in our case), the missing energy in the signal processes is expected to be larger when the jet activity in the background processes is misidentified totally. We have summarized our results for the transverse energy of the jets in comparison with the missing energy in the signal processes in Figure \ref{fig:crossET}. We performed showering and hadronization for the jets by employing Pythia 8 \cite{Bierlich:2022pfr,Reininghaus:2023ctx,Bashir:2023qeb,CMS:2022awf}.}

{The blue dashed curve displays the transverse energy of the jets, while the $E_{T}^{\gamma}$ is represented by the red dashed curve. Since it overlaps with the jet transverse energy it is barely visible. The vertical solid lines show the approximate missing transverse energy of the signal processes for each benchmark point tabled in our manuscript. Note that we have presented the missing energy in the signal events approximately in which $\cancel{E}_{T} \simeq m_{A_{i}}$. In the usual treatment, the missing energy exhibits a distribution around the peak rather than yielding a specific value. The width of the distribution at the half-peak point is determined by the width of the decay mode under concern. In our analyses, the decay width of $A_{i}\rightarrow \tilde{\chi}_{1}^{0} \tilde{\chi}_{1}^{0}$ events are realized as $\Gamma \gtrsim 10^{-5} - 10^{-3}$ GeV. In this case the the missing energy will be distributed around the peak (realized around the mass of CP-odd Higgs bosons) very narrowly, and hence, considering a specific value provides sufficiently precise result. Note that we do not employ this approximation when we consider the missing energy for $Z\gamma$ background events, since the missing energy in these events is formed by $Z\rightarrow \nu\nu$ decays, and the width of the $Z-$boson cannot be neglected here.}

\begin{figure}[h!]
	\subfigure{\includegraphics[scale=0.4]{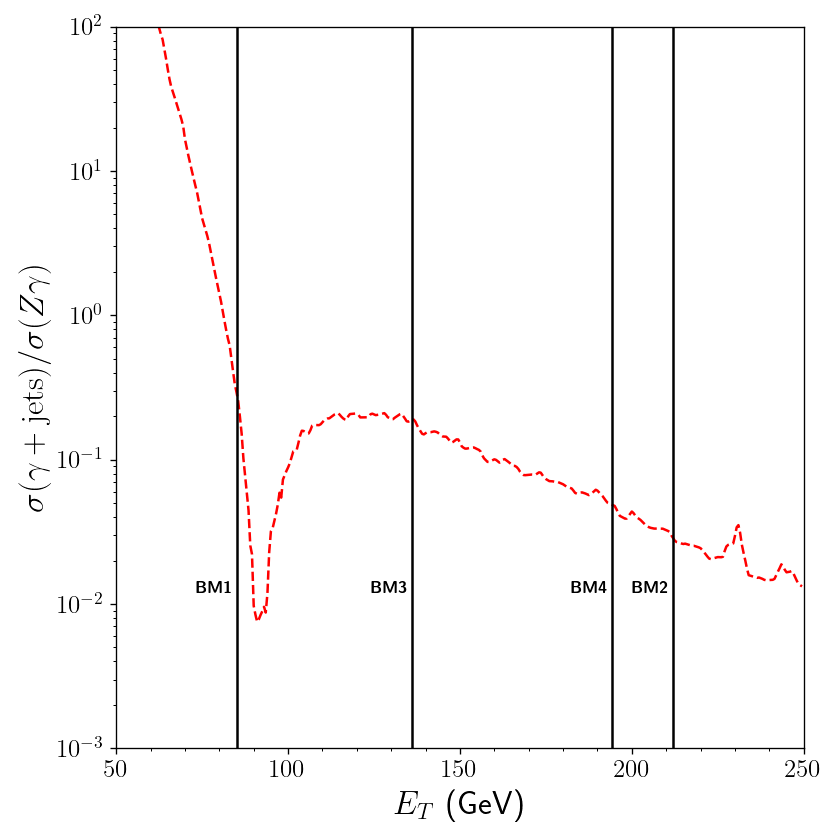}}%
	\subfigure{\includegraphics[scale=0.4]{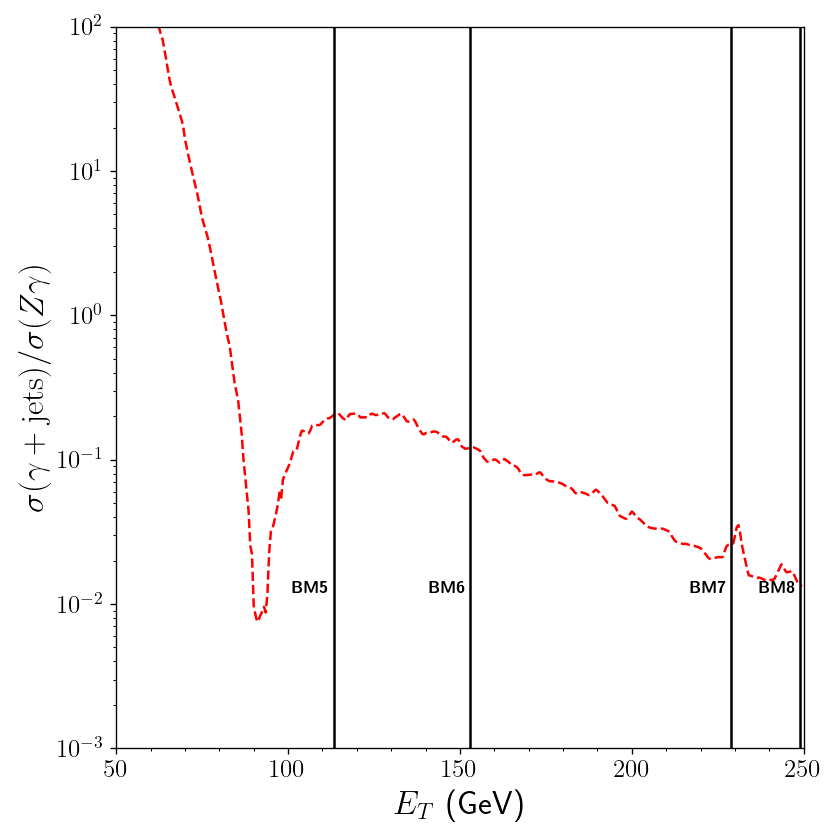}}
	\caption{The ratio of the two sets of the background events formed by the $\gamma+$jets and $Z\gamma$ processes in correlation with the missing energy. In the $\gamma+$jets events, the jets are assumed to be misidentified as missing energy totally. The vertical lines show the missing energies for each benchmark points as pointed in the planes.}
	\label{fig:ratiobcksg}
\end{figure}

{As seen from the plots in Figure \ref{fig:crossET}, the $\gamma +$jets events lose their impact with the increasing transverse energy, and its cross-section drops below around $10^{-2}$ pb when $E_{T} \gtrsim 100$ GeV. If we assume all the jet  activity is misidentified as the missing energy, then these events can imitate the signal processes, but their impact cannot exceed a few percent for relatively heavier CP-odd Higgs boson masses. In this context, BM1, BM3 and BM5 can receive a slightly stronger impact compared to the other benchmark points. However, this impact may not still prevent the visibility of the signal processes represented by these benchmark points. It can be seen by considering the ratios of the cross-sections of $\gamma+$jets events to the $Z\gamma$ processes, which is shown for each benchmark point in Figure \ref{fig:ratiobcksg}. One can enhance the visibility of BM1 (BM3) by applying a cut on the missing energy as $84.5 ~(112.5) \lesssim \cancel{E}_{T} \lesssim 85.5 ~(113.5)$ GeV. Even though these cuts lie below the effective cut ($\cancel{E}_{T} \gtrsim 144.5$ GeV), the $\gamma+$jets events form the background only slightly more than $10\%$ because the $Z\gamma$ yields larger cross-sections in the cases of these benchmark points. Despite being less than the effective cut, the missing energy cut in the case of BM3 ($\simeq 136$ GeV) a cut on the missing energy as $135.5 \lesssim \cancel{E}_{T} \lesssim 136.5$ GeV can still reduce the $\gamma+$jets background to about $10\%$. We should note that the efficiency in identifying the jets is around $50\%$ \cite{CMS:2019gwf}, and so the detector analyses can halve the ratios for the $\gamma+$jets events compared to those which we commented above. In addition, some other cuts might be possible such as those on the angle between the missing energy (or jets) and photon, the pseudo rapidity etc \cite{CMS:2015ifd}. At the level of our analyses, since the photon behaviour is the same for the background and the signal processes, these cuts yield similar effects to the signal and the background processes.}

{Before concluding the possible solutions with a large significance in mono-photon events should be included in the discussions. As mentioned before, the right panel of Figure 6 shows a point with a large significance ($SS_{A_{2}} \simeq 5$), which has been considered to be excluded in our discussions. If one takes a close look, on the other hand, such solutions are allowed through the constrains which we considered in our analyses and included in our discussions. Such solutions emerge when the included $A-$boson is in resonance in mass with the pair of LSP neutralinos of the final states. Indeed, the solutions, which appeared in our analyses, exhibits very similar properties with the fifth benchmark point (BM5 in Table \ref{tab:benchmark2}) in terms of the mixing and masses in the Higgs sector. However, the second lightest CP-odd Higgs boson mass happens to be $m_{A_{2}}\simeq 2m_{\tilde{\chi}_{1}^{0}}$, which enhance the decay width of $A_{2}\rightarrow \tilde{\chi}_{1}^{0} \tilde{\chi}_{1}^{0}$. Such resonance solutions can still be accommodated in our analyses with relatively lower significance when it is sufficiently suppressed by the mixing and/or masses of the Higgs bosons. Indeed, the benchmark points listed in Table \ref{tab:benchmark1}, which are selected for mono-photon events involving $A_{1}$  exemplify also such resonance solutions ($m_{A_{1}}\simeq 2 m_{\tilde{\chi}_{1}^{0}}$). Having only one point for $A_{2}$ in our data sample is due to the statistics of our scans. Since we have considered solutions with such a large significance to be excluded, our scans do not focus on the regions of large significance.}

Finally, we have also displayed the CP-even Higgs boson decays in both sets of the benchmark points. The points in Table \ref{tab:benchmark1} predict cross-sections for the $h_{2}\rightarrow VV$ events about $10^{-3}$, and as discussed in Section \ref{subsec:SMdecays}, these points can also be analyzed through these events in near future. While most of the points in Table \ref{tab:benchmark2} predict lower cross-sections for $h_{2}\rightarrow VV$ events, BM6 will also be included in such analyses, as well, since it yields the largest cross-section for these events realized in our scans. Note that we also assume only one of the singlet CP-odd Higgs bosons can contribute to the mono-photon events in selecting the benchmark points. However, in displaying our results for the significance in Figure 7, we did not consider the efficiency to the missing energy mentioned above. On the other hand, BM1 of Table \ref{tab:benchmark1} and BM5 of Table \ref{tab:benchmark2} suffer from the low efficiency (which is about $30\%$ for BM1 and $55\%$ for BM5 \cite{CMS:2016ljj}), while the other benchmark points approximately correspond to $100\%$ efficiency.

\section{Conclusion}
\label{sec:conc}

We consider a class of Secluded $U(1)^{\prime}-$extended MSSM, in which the models are constrained at $\mgut$ by the universal boundary conditions and family-independent $U(1)^{\prime}$ charges. The secluded sector is spanned by four MSSM singlet scalars. While three of them interact among themselves only, one of the singlets is allowed to have a tree-level coupling with the MSSM Higgs doublets. These models also extend the MSSM particles content by $Z^{\prime}$ - the gauge boson associated with $U(1)^{\prime}$, right-handed neutrinos and exotics. The heavy mass bounds on $Z^{\prime}$ and exotics, and tiny masses experimentally established for the right-handed neutrinos lead these particles to be decoupled from the low scale spectrum, while the singlet scalars can be actively involved in the mass scales from GeV to TeV. In our work, we discuss the implications of the Secluded UMSSM class for the extra scalar states which can be probed in the current and near future collider experiments. We consider models, in which the MSSM fields are also non-trivially charged under $U(1)^{\prime}$, and we identify several sets for the different charge assignments which are compatible with the anomaly cancellations. In exploring the fundamental parameter space, we accept only the solutions which are consistent with the mass bounds, constraints from rare $B-$meson decays and the measurements of the Planck satellite on the relic density of DM within $5\sigma$, and the current results from the direct detection experiments of DM. In this context, the solutions need to yield one of the neutralinos or right-handed sneutrinos to be LSP to be compatible with the DM analyses. In our analyses we consider only the LSP neutralino solutions. We also assume the lightest CP-even Higgs boson is accounted for the SM-like Higgs boson, and we apply the relevant constraints from the current SM-Higgs boson analyses. 

After selecting the solutions allowed by these constraints, we find that the low scale spectra involve CP-even Higgs bosons in the mass interval from about 200 GeV to 2.5 TeV, and two light CP-odd Higgs bosons, whose masses are bounded at about 300 GeV from above  by the DM constraints. The CP-even Higgs bosons can be subjected to the current analyses through their decay modes involving the SM final states such as a pair of $\tau-$leptons, SM gauge bosons or SM-like Higgs bosons. Although these CP-even Higgs bosons yield lower cross-sections for the events involving a pair of $\tau-$leptons or SM-like Higgs bosons, which require more sensitive analyses than those currently performed, they can be probed through their decays into a pair of SM gauge bosons. We find that these solutions can be probed currently up to about $m_{h_{2}} \simeq 2$ TeV, when their mixing with the MSSM Higgs fields are considerably large. Even though smaller mixing yield smaller cross-sections for these events, one can accommodate solutions in the Secluded UMSSM, which are expected to be probed by the analyses in near future. 

While the heavy Higgs bosons can be probed through their decays into SM particles, the light CP-odd Higgs bosons can be involved in a possible signal processes in which they are produced in  associated with a SM particle. In our work, we consider such productions involving a photon in which one of the light CP-odd Higgs boson decays totally into a pair of LSP, while the photon forms a visible final state. Even though the relevant SM background yield a total cross-section of about 10 pb, we show that the SM background can be significantly reduced to about $10^{-3}$ pb with some selections on the missing transverse energy as $\cancel{E}_{T} \gtrsim 100$ GeV and $\cancel{E}_{T} \lesssim 80$ GeV. We find that the mono-photon signals involving the light CP-odd Higgs bosons have cross-section values between $10^{-5}-10^{-3}$ pb. Despite being lower than the SM background, such processes in this cross-section interval can considerably contribute to the total mono-photon events. In this context, the significance can provide a better understanding, and we find that the solutions with the largest cross-section currently yield a significance of 0.8 when $A_{1}$ of a mass around 85 GeV is involved. We display four benchmark points for these processes to exemplify our findings and discuss some possible projections of the near future collider experiments to probe light $A_{1}$ solutions through the mono-photon signals. The selected benchmark points show that the solutions, whose current significance is greater than about 0.5, can be expected to be probed at the end of Run-3 experiments of LHC, while the solutions with significance around 0.3 need to wait for Run-4 experiments. We also identify a lower bound on the cross-section that is about $0.5\times 10^{-5}$ pb. The solutions of mono-photon signals with $A_{1}$ can escape from detection even in HL-LHC experiments, if they yield cross-sections lower than this bound.

We follow similar analyses for the second CP-odd Higgs bosons when its mass lies between 120-230 GeV. Even though its mass is relatively heavier than $A_{1}$, the signals involving $A_{2}$ can be distinguished than the background better than $A_{1}$ in the missing transverse energy plane. The mono-photon signals involving $A_{2}$ yield similar cross-section values as those involving $A_{1}$, but since the background is significantly lowered, the solutions with $m_{A_{2}}\simeq 120$ can be excluded already, while a smaller mixing between the MSSM Higgs fields can leave some solutions in this region available for the upcoming analyses. We also represent four benchmark points to exemplify the significance of these processes to measure their contributions to the total mono-photon events. The solutions with a significance about 0.7 are expected to be probed at the end of Run-3 experiments. In contrast to the solutions with $A_{1}$, the signal processes involving $A_{2}$ can potentially be analyzed in Run-4 and HL-LHC experiments even if they have about $0.2-0.3$ significance in the current experiments. 

\section*{Acknowledgment}

CSU would like to thank Instituto de F\'{i}sica The\'{o}rica of Universidad Aut\'{o}noma de Madrid, where part of his research has been conducted. The work of YH is supported by  Balikesir University Scientific Research Projects with Grant No. BAP-2022/083.

\bibliographystyle{JHEP}
\bibliography{Secluded}

\end{document}